\documentclass[twocolumn]{emulateapj}  
\usepackage{times}

\usepackage{graphicx,bm,amssymb}
\usepackage{epsfig}
\usepackage{amssymb}
\usepackage{natbib}
\usepackage{pstricks}
\usepackage{psfrag}
\usepackage[colorlinks=true,linkcolor=blue,citecolor=blue]{hyperref}
\usepackage{float}
\usepackage[caption = false]{subfig}
\usepackage{rotating}
\def\araa{ARA\&A}
\def\apj{ApJ}
\def\apjl{ApJ}
\def\apjs{ApJS}
\def\aap{A\&A}
\def\mnras{MNRAS}
\def\prl{Phys.~Rev.~Lett.}
\def\nat{Nature}
\def\physrep{Phys.~Rep.}

\newcommand{\be}{\begin{equation}}
\newcommand{\ee}{\end{equation}}
\newcommand{\bary}{\begin{eqnarray}}
\newcommand{\eary}{\end{eqnarray}}

\shorttitle{Atypical bursts}
\shortauthors{Fraija N.}

\begin{document}
\title{Description of atypical bursts seen slightly off-axis}
\author{N. Fraija$^{1\dagger}$, F.  De Colle$^2$, P. Veres$^3$, S.~ Dichiara$^{4,5}$, R. Barniol Duran$^6$, A.C. Caligula do E. S. Pedreira$^{1}$,  A. Galvan-Gamez$^{1}$ and B. Betancourt Kamenetskaia$^{1}$}
\affil{$^1$ Instituto de Astronom\' ia, Universidad Nacional Aut\'onoma de M\'exico, Circuito Exterior, C.U., A. Postal 70-264, 04510 Cd. de M\'exico,  M\'exico.\\
$^2$ Instituto de Ciencias Nucleares, Universidad Nacional Aut\'onoma de M\'exico, Circuito Exterior, C.U., A. Postal 70-264, 04510 Cd. de M\'exico,  M\'exico\\
$^3$ Center for Space Plasma and Aeronomic Research (CSPAR), University of Alabama in Huntsville, Huntsville, AL 35899, USA\\
$^4$ Department of Astronomy, University of Maryland, College Park, MD 20742-4111, USA\\
$^5$ Astrophysics Science Division, NASA Goddard Space Flight Center, 8800 Greenbelt Rd, Greenbelt, MD 20771, USA\\
$^6$ Department of Physics and Astronomy, California State University, Sacramento, 6000 J Street, Sacramento, CA 95819-6041, USA\\
}
\email{$\dagger$nifraija@astro.unam.mx}
\date{\today} 
%
\begin{abstract}
%
The detection of gravitational waves together with their electromagnetic counterpart,  in the gamma-ray burst GRB 170817A,  marked a new era of multi-messenger astronomy. Several theoretical models have been proposed to explain the atypical behaviour of this event.   Recently,  it was shown that the multi-wavelength afterglow of GRB 170817A  was consistent with  a synchrotron forward-shock model when the outflow was viewed off-axis,  decelerated in a uniform medium and parametrized through  a power-law velocity distribution.    Motivated by  the upper limits on the very-high-energy emission,  and  the stratified  medium in the  close vicinity of a binary neutron star merger proposed to explain the gamma-ray flux in the short GRB 150101B, we extend the mechanism proposed to explain GRB 170817A to a more general scenario deriving the synchrotron self-Compton (SSC) and synchrotron forward-shock model when the off-axis outflow is decelerated in a uniform and stratified circumburst density.   As particular cases,  we show that the delayed and long-lasting afterglow emission observed in  GRB 080503,  GRB 140903A, GRB 150101B and GRB 160821B could be interpreted by a similar scenario to the one used to describe GRB 170817A.  In addition, we show that the proposed scenario agrees with the MAGIC, Fermi-LAT and H.E.S.S upper limits on gamma-ray emission from GRB 160821B and GRB 170817A. 
\end{abstract}
\section{Introduction}
%
%
Gravitational wave (GW) detection with its electromagnetic counterpart marked a new era of multi-messenger astronomy.  The second run (02) of the Advanced Laser Interferometer Gravitational-Wave Observatory (LIGO) and Advanced Virgo \citep{PhysRevLett.119.161101,2041-8205-848-2-L12} led to the important discovery of the first GWs associated to the short gamma-ray burst GRB 170817A which was detected by Gamma-ray Burst Monitor (GBM) onboard Fermi Gamma-ray Space Telescope  \citep{2017ApJ...848L..14G} and The INTErnational Gamma-Ray Astrophysics Laboratory \citep[INTEGRAL;][]{2017ApJ...848L..15S}. The progenitor of this transient event was promptly associated to the merger of two neutron stars (NSs) located in the host galaxy NGC 4993, at a redshift of $z\simeq0.01$.   \citep{2017Sci...358.1556C,2017ApJ...848L..20M}. Due to its low luminosity and the detection of  a delayed and long-lasting non-thermal emission (afterglow) observed in radio, optical and  X-ray bands, this short gamma-ray burst (sGRB) was classified as atypical  \citep{2017Natur.547..425T, 2017ApJ...848L..20M, 2018arXiv180103531M, 2017ATel11037....1M, 2018ATel11242....1H, 2019ApJ...870L..15L}. These long-lasting observations were described by synchrotron emission generated by the deceleration of  off-axis top-hat jets \citep{2017Natur.551...71T,2017ApJ...848L..20M, 2017arXiv171005905I, 2017ApJ...848L..21A, 2019ApJ...871..123F, 2019arXiv190210303G}, radially stratified ejecta \citep{2017arXiv171111573M, 2019ApJ...871..200F,2018ApJ...867...95H} and structured jets \citep{2017Sci...358.1559K, 2017MNRAS.472.4953L, 2017arXiv171203237L,2019MNRAS.484L..98K} in a homogeneous medium.   In particular,   it was shown in \cite{2019ApJ...871..123F} that the delayed non-thermal multi-wavelength emission observed in GRB 170817A was consistent with  the synchrotron forward-shock model when the outflow was viewed off-axis,  decelerated in a homogeneous medium and parametrized through  a power-law velocity distribution. \\
%
%
Similar observational features of GRB 170817A such as a short gamma-ray spike and an undetected afterglow in a timescale of days followed by a very bright emission in X-rays, optical and/or radio bands can support the idea that sGRBs generally launch collimated outflows out of the observer's line of sight.  This is the case of the short GRB 080503 \citep{2009ApJ...696.1871P},   GRB 140903A \citep{2016ApJ...827..102T},  GRB 150101B \citep{2016ApJ...833..151F, 2018NatCo...9.4089T, 2018ApJ...863L..34B} and GRB 160821B \citep{2017ApJ...835..181L, 2016GCN.19843....1S, 2019MNRAS.489.2104T, 2019ApJ...883...48L, 2018ApJ...857..128J} that exhibited a short gamma-ray spike together with a rebrightening in a timescale of hours to days detected in several energy bands.  The GBM Collaboration studied GRB 150101B and found that the gamma-ray light curve composed of a short hard spike and a long soft tail exhibited similar features to those of GRB 170817A \citep{2018ApJ...863L..34B}.  This collaboration  derived the condition for the long tail occurring at the external shocks in a stratified stellar-wind like medium.\\
%
%
Since the Fermi satellite began scientific operations, the Large Area Telescope (LAT) has reported the detection of very-high-energy photons (VHE; $\gtrsim 10$ GeV)  in more than a dozen GRBs \citep[see][and references therein]{2009ApJ...706L.138A, 2013ApJS..209...11A,2014Sci...343...42A, 2014ApJ...787L...6L, 2016GCN.19413....1L, 2017ApJ...848...15F, 2019ApJ...885...29F, 2019ApJ...883..162F, 2019ApJ...879L..26F}.  Although the search for VHE photons by means of Imaging Atmospheric Cherenkov Telescopes (IACTs) has been a challenge  because the time required to re-point to the burst position may take minutes,   the MAGIC telescope, recently, detected photons in the direction of GRB 190114C  with energies above 300 GeV for almost 20 minutes \citep{2019GCN.23701....1M}.     In the framework of the fireball model, the standard synchrotron radiation originated during the deceleration phase has been successful in explaining the long-lasting emission.  However, this is not the case  when the photons detected are greater than the maximum synchrotron photon energy  $\sim 10~{\rm GeV}~\left(\frac{\Gamma}{200}\right)\left(\frac{1+z}{2}\right)^{-1}$ \citep[][and references therein]{2010ApJ...718L..63P, 2011MNRAS.412..522B,  2020arXiv200311252F}, where $\Gamma$ is the bulk Lorentz factor of the decelerated outflow.  In order to interpret the VHE photons, the standard synchrotron self-Compton (SSC) model in the forward shocks has been used \citep[e.g. see,][]{2001ApJ...548..787S}.\\
In this paper, the mechanism proposed to describe GRB 170817A and introduced in \cite{2019ApJ...871..123F} is extended to a more general scenario deriving the synchrotron and synchrotron self-Compton (SSC) emission from  forward shocks when the outflow, parametrized with a power-law velocity distribution, is decelerated in a homogeneous ISM-like and in a stellar wind-like medium.  We show that the delayed non-thermal emission observed in GRB 080503, GRB 140903A, GRB 150101B and GRB 160821B could be interpreted by
a jet with a velocity distribution seen slightly off-axis.   In addition, we show that the proposed scenario agrees with the VHE gamma-ray upper limits derived by the GeV - TeV observatories.   The paper is organized as follows: In Section 2  we present the SSC and synchrotron forward-shock model when the outflow is decelerated in a homogeneous and in a wind-like medium. In Section 3 we apply this model to describe the delayed multi-wavelength afterglow  observed in GRB 080503, GRB 140903A, GRB GRB 150101B and GRB 160821B, and also to obtain the SSC emission that should have been emitted during GRB 160821B and GRB 170817A.  In Section 4,  we present our conclusions.\\
%
\section{Theoretical model}
Once the outflow launched by the NS merger sweeps up enough  circumburst material (stratified wind-like  and/or uniform ISM-like medium),  electrons originally accelerated during the forward shocks are cooled down by synchrotron and SSC radiation.   We use the corresponding equivalent kinetic energy defined in \cite{2019ApJ...871..123F} 
\bary\label{Eoff}
E_{\rm k}&=& \tilde{E}\,\Gamma^{-\alpha_s} (1+ \Delta \theta^2\Gamma^2)^{-3}\,,
\eary
with $\tilde{E}$ the fiducial energy, and $\Delta\theta= \theta_{\rm obs} - \theta_j$ defined by the viewing angle ($\theta_{\rm obs}$)  and the opening angle ($\theta_j$).   The kinetic energy can be interpreted as the contribution of two parts: i) An off-axis jet concentrated within an opening angle (``top-hat jet") with equivalent kinetic energy $\propto (1+ \Delta \theta^2\Gamma^2)^{-3}$  and ii) an isotropic material with equivalent kinetic energy $\propto \Gamma^{-\alpha_s}$ with $\alpha_s=1.1$ for $\beta\Gamma\gg 1$ and $\alpha_s=5.2$ for $\beta\Gamma\ll 1$ for the adiabatic case  \citep{2001ApJ...551..946T, 2000ApJ...535L..33S, 2015MNRAS.448..417B, 2015MNRAS.450.1430H, 2013ApJ...778L..16H, 2014MNRAS.437L...6K,  2019ApJ...871..200F}.\\
%
%
\subsection{Uniform ISM-like Medium}
\subsubsection{Relativistic stage}
In the relativistic  regime ($\Gamma^2\Delta \theta^2\gg1$),  the equivalent kinetic energy becomes $E_{\rm k}= \tilde{E}\,\Delta \theta^{-6} \Gamma^{-\delta}$ with $\delta=\alpha_s+6$. Given the adiabatic evolution of the forward shock \citep{1976PhFl...19.1130B, 1997ApJ...489L..37S}, the bulk Lorentz  factor evolves  as 
\be\label{Gamma}
\Gamma=  10.1\,\left(\frac{1+z}{1.022}\right)^{\frac{3}{\delta+8}}\, n^{-\frac{1}{\delta+8}}_{-4} \, \Delta \theta^{-\frac{6}{\delta+8}}_{15^\circ} \,  \tilde{E}^{\frac{1}{\delta+8}}_{52}  \,t^{-\frac{3}{\delta+8}}_{1\,{\rm d}}\,,
\ee
where the fiducial energy is given by $\tilde{E}=  \frac{32\pi}{3}\,m_p  (1+z)^{-3}\,n\,\Gamma^{\delta+8}\,\Delta \theta^6\, t^3$ with $m_p$ the proton mass, $z$ the redshift,  $n$ the number density of the uniform ISM-like medium and $t$ the timescale of the outflow during the deceleration phase.  An hypothetical event located at 100 Mpc ($z\approx 0.022$) is considered.   The convention $\hbar=c=1$ in natural units,  $Q_{\rm x}=Q/10^{\rm x}$ in c.g.s. units and the values of cosmological parameters reported in \cite{2018arXiv180706209P}  are adopted.\\ 
\paragraph{Synchrotron Light Curves.} Using the bulk Lorentz factor (eq. \ref{Gamma})  and the synchrotron afterglow theory introduced in \cite{1998ApJ...497L..17S} for the fully adiabatic regime, we derive, in this formalism,  the relevant quantities of synchrotron emission originated from the  forward shocks. The minimum and cooling electron Lorentz factors are  given by
{\small
\bary\label{eLor_syn_ism}
\gamma_m&=& 33.6\,\left(\frac{1+z}{1.022}\right)^{\frac{3}{\delta+8}}\, g(p)\, \epsilon_{e,-2} \, n_{-4}^{-\frac{1}{\delta+8}}\,\Delta\theta^{-\frac{6}{\delta+8}}_{15^\circ}\, \tilde{E}^{\frac{1}{\delta+8}}_{52}\cr
&&\hspace{5.8cm} \times \,t^{-\frac{3}{\delta+8}}_{1\,{\rm d}}\cr
\gamma_c&=& 4.3\times 10^8 \left(\frac{1+z}{1.022}\right)^{\frac{\delta-1}{\delta+8}} (1+Y)^{-1}\, \epsilon_{B,-4}^{-1}\,n_{-4}^{-\frac{\delta+5}{\delta+8}}\,\Delta\theta^{\frac{18}{\delta+8}}_{15^\circ} \,\cr
&&\hspace{4.8cm} \times\tilde{E}^{-\frac{3}{\delta+8}}_{52}\,t^{\frac{1-\delta}{\delta+8}}_{1\,{\rm d}}\,,
\eary
}
respectively, which correspond to a comoving magnetic field given by
%
%
{\small
\bary
B'&\simeq&0.4\,\,{\rm mG}\,\left(\frac{1+z}{1.022}\right)^{\frac{3}{\delta+8}}\, \epsilon^{\frac12}_{B,-4}\, n_{-4}^{\frac{\delta+6}{2(\delta+8)}}\, \Delta\theta^{-\frac{6}{\delta+8}}_{15^\circ}\cr
&&\hspace{5cm} \times \,E^{\frac{1}{\delta+8}}_{52}\,t_{1\,{\rm d}}^{-\frac{3}{\delta+8}}\,.
\eary
}
%
Here,  $Y$ is the Compton parameter, $g(p)=(p-2)/(p-1)$ with $p$ the  spectral index of the electron population,  $\epsilon_e$ and $\epsilon_B$ are the microphysical parameters related to the energy density given to accelerate electrons and amplify the magnetic field, respectively.  Using the electron Lorentz factors (eq. \ref{eLor_syn_ism}),  the characteristic and cooling spectral breaks for synchrotron radiation are
{\small
\bary\label{En_br_syn_ism}
\epsilon^{\rm syn}_{\rm m}&\simeq& 2.3\times 10^{-3}\,{\rm GHz}\,\, \left(\frac{1+z}{1.022}\right)^{-\frac{\delta - 4}{\delta+8}}\, \epsilon^2_{e,-2}\,\epsilon_{B,-4}^{\frac12}\,n_{-4}^{-\frac{\delta}{2(\delta+8)}}\cr
&&\hspace{4cm} \times\,\Delta\theta_{15^\circ}^{-\frac{24}{\delta+8}}\,E_{52}^{\frac{4}{\delta+8}}\,  t_{1\,{\rm d}}^{-\frac{18}{\delta+8}}\cr
\epsilon^{\rm syn}_{\rm c}&\simeq&  8.7\times 10^{4}\,{\rm keV}\,\, \left(\frac{1+z}{1.022}\right)^{\frac{\delta - 4}{\delta+8}}\, (1+Y)^{-2} \,\epsilon_{B,-4}^{-\frac32}\,n_{-4}^{-\frac{3\delta+16}{2(\delta+8)}}\,\cr
&&\hspace{3.6cm} \times\,  \Delta\theta_{15^\circ}^{\frac{24}{\delta+8}}\, E_{52}^{-\frac{4}{\delta+8}}\,  t_{1\,{\rm d}}^{-\frac{2(\delta+2)}{\delta+8}}\,,
\eary
}
respectively.  Considering the maximum emissivity, the total number of radiating electrons and the luminosity distance $D$ from this hypothetical event,  the maximum flux emitted by synchrotron radiation is given by
{\small
\bary\label{flux_syn}
F^{\rm syn}_{\rm max} &\simeq& 1.6\times 10^{-1}\,{\rm mJy}\,\, \left(\frac{1+z}{1.022}\right)^{\frac{16-\delta}{\delta+8}}\,\epsilon_{B,-4}^{\frac12}\,n_{-4}^{\frac{3\delta+8}{2(\delta+8)}}\,\Delta\theta_{15^\circ}^{-\frac{48}{\delta+8}}\cr
&&\hspace{4.4cm} \times\, D^{-2}_{26.3}\,E_{52}^{\frac{8}{\delta+8}}\,  t_{1\,{\rm d}}^{\frac{3\delta}{\delta+8}}\,.
\eary
}
Using the spectral breaks (eq. \ref{En_br_syn_ism})   and the maximum flux (eq. \ref{flux_syn}), the light curves of the synchrotron emission evolving  in the fast- and slow-cooling regime can be written  as
{\small 
\begin{eqnarray}\label{Fl_syn_ism_fc}
\label{scsyn_t}
F^{\rm syn}_{\nu}\propto\cases{
t^{\frac{11\delta+4}{3(\delta+8)}}  \, \epsilon_\gamma^{\frac13},\hspace{1.3cm} \epsilon_\gamma<\epsilon^{\rm syn}_{\rm c}\cr
t^{\frac{2(\delta-1)}{\delta+8}}\,\epsilon_\gamma^{-\frac{1}{2}}\,\hspace{1.2cm}  \epsilon^{\rm syn}_{\rm c}<\epsilon_\gamma<\epsilon^{\rm syn}_{\rm m}\,\,\,\,\,\cr
t^{\frac{2\delta-6p+4}{\delta+8}}\,\epsilon_\gamma^{-\frac{p}{2}},\hspace{0.85cm}\epsilon^{\rm syn}_{\rm m}<\epsilon_\gamma\,, \cr
}
\end{eqnarray}
}
and
{\small
\begin{eqnarray}\label{Fl_syn_ism_sc}
F^{\rm syn}_{\nu}\propto\cases{
t^{\frac{3\delta+4}{\delta+8}}  \, \epsilon_\gamma^{\frac13},\hspace{1.3cm} \epsilon_\gamma<\epsilon^{\rm syn}_{\rm m}\cr
t^{\frac{3\delta-6p+6}{\delta+8}}\,\epsilon_\gamma^{-\frac{p-1}{2}},\hspace{0.3cm} \epsilon^{\rm syn}_{\rm m}<\epsilon_\gamma<\epsilon^{\rm syn}_{\rm c}\,\,\,\,\,\cr
t^{\frac{2\delta-6p+4}{\delta+8}}\,\epsilon_\gamma^{-\frac{p}{2}},\hspace{0.6cm}\epsilon^{\rm syn}_{\rm c}<\epsilon_\gamma\,, \cr
}
\end{eqnarray}
}
respectively.  Considering the particular scenario of $\delta=0$, the observable quantities derived in \cite{1998ApJ...497L..17S} and the light curves of the synchrotron forward-shock emission are recovered \citep[e.g., see][]{2016ApJ...831...22F}.\\ 
\paragraph{SSC Light Curves}  Synchrotron photons generated at the forward shock can be up-scattered by the same electron population as $\epsilon^{\rm ssc}_{\rm m}\sim\gamma^2_{\rm m} \epsilon^{\rm syn}_{\rm m}$ and  $\epsilon^{\rm ssc}_{\rm c}\sim\gamma^2_{\rm c} \epsilon^{\rm syn}_{\rm c}$ \cite[e.g, see][]{2014MNRAS.437.2187F}.  Therefore,  given eqs. (\ref{eLor_syn_ism}) and (\ref{En_br_syn_ism}), the characteristic and cooling spectral breaks for the SSC process  in the fully adiabatic regime are
{\small
\bary\label{En_br_ssc_ism}
\epsilon^{\rm ssc}_{\rm m}&\simeq& 1.1\times 10^{-5}\,{\rm eV}\,\, \left(\frac{1+z}{1.022}\right)^{\frac{10 - \delta}{\delta+8}}\, g(p)^2\,\,\epsilon^4_{e,-2}\,\epsilon_{B,-4}^{\frac12}\,n_{-4}^{\frac{\delta-4}{2(\delta+8)}}\,\cr
&&\hspace{4.7cm} \times\,\Delta\theta_{15^\circ}^{-\frac{36}{\delta+8}} \,E_{52}^{\frac{6}{\delta+8}}\,  t_{1\,{\rm d}}^{-\frac{18}{\delta+8}}\cr
\epsilon^{\rm ssc}_{\rm c}&\simeq&  1.6\times 10^{13}\,{\rm TeV}\,\, \left(\frac{1+z}{1.022}\right)^{\frac{3(\delta - 2)}{\delta+8}}\,(1+Y)^{-4} \,\epsilon_{B,-4}^{-\frac72}\,n_{-4}^{-\frac{7\delta+36}{\delta+8}}\cr
&&\hspace{3.5cm} \times\,\Delta\theta_{15^\circ}^{\frac{60}{\delta+8}} \,E_{52}^{-\frac{10}{\delta+8}}\,  t_{1\,{\rm d}}^{-\frac{2(2\delta+1)}{\delta+8}}\,,
\eary
}
respectively.  The break energy, due to  the Klein-Nishina (KN) effect,  is given by
{\small
\bary
\epsilon^{\rm ssc}_{\rm KN}&\simeq& 2.4\times10^3 \,{\rm GeV}\,\, \left(\frac{1+z}{1.022}\right)^{-\frac{6}{\delta+8}}\, (1+Y)^{-1} \,\epsilon_{B,-4}^{-1}\,n_{-4}^{-\frac{\delta+6}{\delta+8}}\,\cr
&&\hspace{3.2cm} \times\,  \Delta\theta_{15^\circ}^{\frac{12}{\delta+8}}\, E_{52}^{-\frac{2}{\delta+8}}\,  t_{1\,{\rm d}}^{-\frac{\delta+2}{\delta+8}}\,.
\eary
}
Considering the maximum flux of the synchrotron radiation and the optical depth \citep[see][]{2001ApJ...548..787S}, the maximum flux emitted by the SSC process is given by
{\small
\bary\label{flux_ssc}
F^{\rm ssc}_{\rm max} &\simeq& 5.1\times 10^{-11}\,{\rm mJy} \left(\frac{1+z}{1.022}\right)^{-\frac{5(\delta+2)}{\delta+8}}\epsilon_{B,-4}^{\frac12}\,n_{-4}^{\frac{5(\delta+4)}{2(\delta+8)}}\Delta\theta_{15^\circ}^{-\frac{60}{\delta+8}}\cr
&&\hspace{4.1cm} \times\, D^{-2}_{26.3} \,E_{52}^{\frac{10}{\delta+8}}\,  t_{1\,{\rm d}}^{\frac{2(2\delta+1)}{\delta+8}}.
\eary
}
Using the spectral breaks (eq. \ref{En_br_ssc_ism})  and the maximum flux (eq. \ref{flux_ssc}), the light curves of the SSC process evolving  in the fast- and slow-cooling regime can be written  as
{\small
\begin{eqnarray}\label{Fl_ssc_ism_fc}
F^{\rm ssc}_{\nu}\propto\cases{
t^{\frac{8(2\delta+1)}{3(\delta+8)}}  \, \epsilon_\gamma^{\frac13},\hspace{1.3cm} \epsilon_\gamma<\epsilon^{\rm ssc}_{\rm c},\cr
t^{\frac{2\delta+1}{\delta+8}}\,\epsilon_\gamma^{-\frac{1}{2}},\hspace{1.3cm} \epsilon^{\rm ssc}_{\rm c}<\epsilon_\gamma<\epsilon^{\rm ssc}_{\rm m},\,\,\,\,\,\cr
t^{\frac{10+2\delta-9p}{\delta+8}}\,\epsilon_\gamma^{-\frac{p}{2}},\hspace{0.9cm} \epsilon^{\rm ssc}_{\rm m}<\epsilon_\gamma\,, \cr
}
\end{eqnarray}
}
and
{\small
\begin{eqnarray}\label{Fl_ssc_ism_sc}
F^{\rm ssc}_{\nu}\propto \cases{
t^{\frac{4(\delta+2)}{\delta+8}}  \, \epsilon_\gamma^{\frac13},\hspace{1.3cm} \epsilon_\gamma<\epsilon^{\rm ssc}_{\rm m},\cr
t^{\frac{4\delta-9p+11}{\delta+8}}\,\epsilon_\gamma^{-\frac{p-1}{2}},\hspace{0.4cm} \epsilon^{\rm ssc}_{\rm m} <\epsilon_\gamma<\epsilon^{\rm ssc}_{\rm c},\,\,\,\,\,\cr
t^{\frac{2\delta-9p+10}{\delta+8}}\,\epsilon_\gamma^{-\frac{p}{2}},\hspace{0.7cm} \epsilon^{\rm ssc}_{\rm c}<\epsilon_\gamma\,, \cr
}
\end{eqnarray}
}
respectively. Considering the particular scenario of $\delta=0$, the observable quantities derived in \cite{2001ApJ...548..787S}  are recovered.\\ 
\subsubsection{Lateral expansion stage}
\begin{figure}[h!]
{ \centering
\resizebox*{1.\textwidth}{0.6\textheight}
{\includegraphics{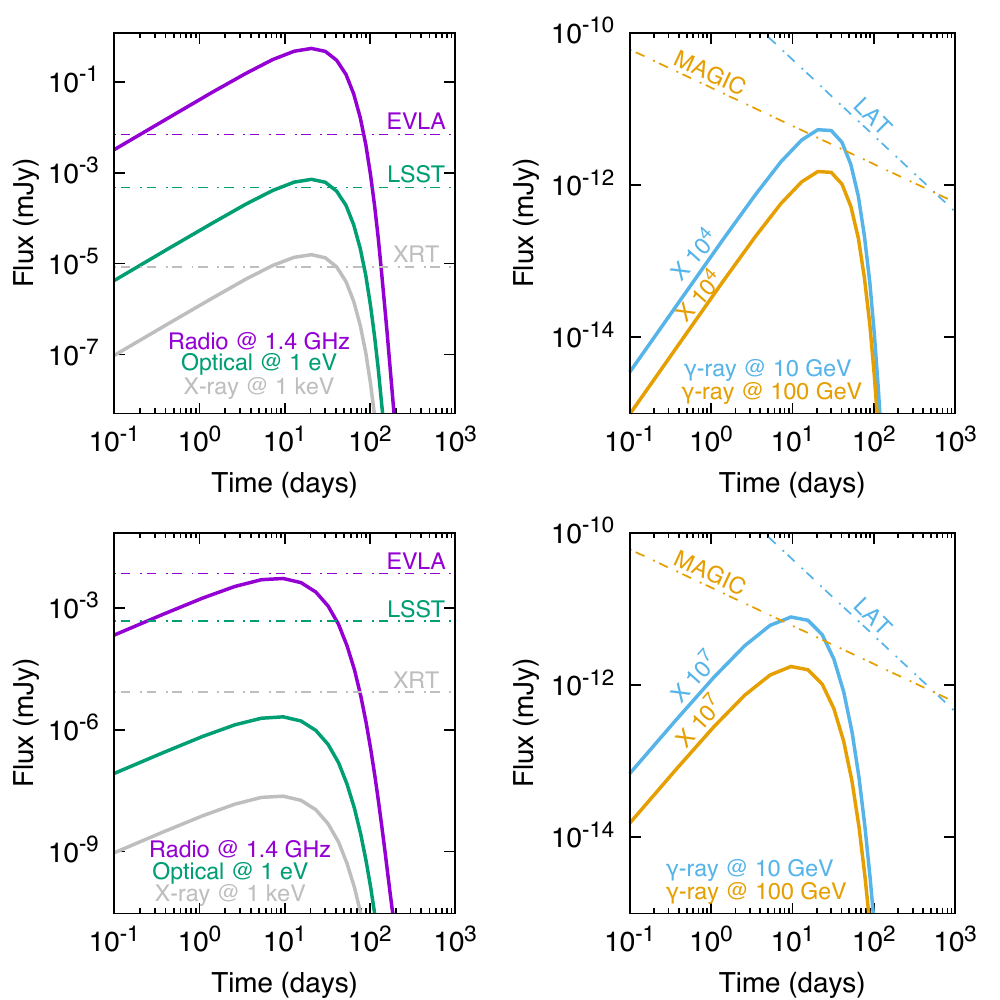}}
}
\caption{The figures show the light curves of the synchrotron (left-hand panels) and SSC (right-hand panels) fluxes radiated from the decelerated outflow in a homogeneous density for the values of $\tilde{E}=5\times 10^{52}\,{\rm erg}$, $n=10^{-4}\,{\rm cm^{-3}}$,  $\epsilon_B=10^{-2}$, $\Delta \theta=20^\circ$ and $\alpha_s=2.1$ (upper panels) and  $\tilde{E}=10^{51}\,{\rm erg}$,  $n=10^{-3}\,{\rm cm^{-3}}$, $\epsilon_B=10^{-4}$,  $\Delta \theta=30^\circ$ and $\alpha_s=1.1$ (lower panels). The values of $\epsilon_e=0.1$, $p=2.2$ and $D=100\, {\rm Mpc}$ were assumed in all the panels.}
 \label{fig1:jet_ism}
\end{figure}

During the lateral expansion stage, the beaming cone of the radiation emitted off-axis broadens increasingly until this cone reaches the observer's field of view \citep[$\Gamma\sim \Delta \theta^{-1}$; ][]{2002ApJ...570L..61G,  2017arXiv171006421G}.   Recently, based on relativistic numerical jet calculations during this stage, \cite{2018ApJ...865...94D} presented a semi-analytical model to calculate the corresponding Lorentz factor and opening angle as the jet spreads; however, here we treat this stage approximately \citep[e.g.,][]{2002ApJ...570L..61G}.   Given that the timescale for the lateral expansion phase to occur is much longer than the timescale of the transition from the fast- to the slow-cooling regime, only the synchrotron and SSC light curves in the slow-cooling regime are derived in this stage.  Given the Blanford-McKee solution and the equivalent kinetic energy (eq. \ref{Eoff}), during the lateral expansion stage the kinetic energy can be approximated as $E_{\rm k}\approx \frac18 \tilde{E}\,\Gamma^{-\alpha_s}$. In this approximation,   the bulk Lorentz factor evolves as
\be\label{Gamma_l_ism}
\Gamma=  4.1\left(\frac{1+z}{1.022}\right)^{\frac{3}{\delta+8}}\, n^{-\frac{1}{\delta+8}}_{-4}\, \Delta \theta^{-\frac{6}{\delta+8}}_{15^\circ} \,  \tilde{E}^{\frac{1}{\delta+8}}_{52}  \,t^{-\frac{3}{\delta+8}}_{10\,{\rm d}}\,.
\ee
Similarly, the timescale for the cone to reach the observer's field of view can be written as \citep{2017arXiv171006421G}
{\small
\be\label{lat_exp_jt}
t_{\rm br}=23.3\,{\rm d}\,\,{\rm k} \,\left(\frac{1+z}{1.022}\right)\, n^{-\frac{1}{3}}_{-4}\, \tilde{E}_{52}^{\frac{1}{3}}\,  \Delta \theta_{15^\circ}^2\,,
\ee
}
where the value of the parameter ${\rm k}$ varies from one model to another \citep{2002ApJ...579..699N, 2002ApJ...570L..61G}.   In this case,  the fiducial energy can be obtained from eq. (\ref{Gamma_l_ism}).  
%
%
%
\paragraph{Synchrotron Light Curves.} Using the bulk Lorentz factor (eq. \ref{Gamma_l_ism}) and the evolution of a jet after it slows down and spreads laterally  introduced in \cite{1999ApJ...519L..17S}, we derive the electron Lorentz factors, the spectral breaks, the maximum flux and the light curves when the synchrotron emission is evolving  in the fully adiabatic slow-cooling regime. The minimum and cooling electron Lorentz factors are  given by
{\small
\bary\label{eLor_syn_ism_l}
\gamma_{\rm m}&=& 14.4\left(\frac{1+z}{1.022}\right)^{\frac{3}{\delta+6}}\, g(p)\,\epsilon_{\rm e, -2}\, n_{-4}^{-\frac{1}{\delta+6}}\, \tilde{E}^{\frac{1}{\delta+6}}_{52}\,t^{-\frac{3}{\delta+6}}_{10\,{\rm d}}\cr
\gamma_{\rm c}&=& 2.7\times 10^8 \left(\frac{1+z}{1.022}\right)^{\frac{\delta-3}{\delta+6}} (1+Y)^{-1}\, \epsilon_{B,-4}^{-1}\,n_{-4}^{-\frac{\delta+3}{\delta+6}}\,\cr
&&\hspace{4.5cm} \times\,\tilde{E}^{-\frac{3}{\delta+6}}_{52}\,t^{\frac{3-\delta}{\delta+6}}_{10\,{\rm d}}\,,
\eary
}
respectively. Their corresponding  characteristic and cooling spectral breaks become
{\small
\bary\label{En_br_syn_ism_l}
\epsilon^{\rm syn}_{\rm m}&\simeq& 7.8\times 10^{-5}\,{\rm GHz}\,\, \left(\frac{1+z}{1.022}\right)^{\frac{6-\delta }{\delta+6}}\, g(p) \, \epsilon^2_{e,-2}\,\epsilon_{B,-4}^{\frac12}\,n_{-4}^{\frac{\delta-2}{2(\delta+6)}}\,\cr
&&\hspace{5cm} \times\,E_{52}^{\frac{4}{\delta+6}}\,  t_{10\,{\rm d}}^{-\frac{12}{\delta+6}}\cr
\epsilon^{\rm syn}_{\rm c}&\simeq&  8.7\times 10^{3}\,{\rm keV}\,\, \left(\frac{1+z}{1.022}\right)^{\frac{\delta - 6}{\delta+6}}\,(1+Y)^{-2}\, \epsilon_{B,-4}^{-\frac32}\,n_{-4}^{-\frac{3\delta+10}{2(\delta+6)}}\,\cr
&&\hspace{4.5cm} \times E_{52}^{-\frac{4}{\delta+6}}\,  t_{10\,{\rm d}}^{-\frac{2\delta}{\delta+6}}\,.
\eary
}
Taking into account the maximum emissivity, the total number of radiating electrons and the distance from this source,  the maximum flux radiated by synchrotron emission during this phase is given by
{\small
\bary\label{flux_later}
F^{\rm syn}_{\rm max} &\simeq& 7.2\times 10^{-13}\,{\rm mJy}\,\, \left(\frac{1+z}{1.022}\right)^{\frac{18-\delta}{\delta+6}}\,\epsilon_{B,-4}^{\frac12}\,n_{-4}^{\frac{3\delta+2}{2(\delta+6)}}\,E_{52}^{\frac{8}{\delta+6}}\,\cr
&&\hspace{4.5cm} \times  D^{-2}_{26.3}\,t_{10\,{\rm d}}^{-\frac{3(2 - \delta)}{\delta+6}}\,.
\eary
}
Using the spectral breaks (eq. \ref{En_br_syn_ism_l})  and the maximum flux (eq. \ref{flux_later}), then the light curves of the synchrotron emission evolving  in the slow-cooling regime become
{\small
\begin{eqnarray}\label{Fl_syn_ism_sc_l}
F^{\rm syn}_{\nu}\propto\cases{
t^{\frac{3\delta-2}{\delta+6}}  \, \epsilon_\gamma^{\frac13},\hspace{1.3cm} \epsilon_\gamma<\epsilon_{\rm m},\cr
t^{\frac{3(\delta-2p)}{\delta+6}}\,\epsilon_\gamma^{-\frac{p-1}{2}},  \hspace{0.5cm}   \epsilon_{\rm m}<\epsilon_\gamma<\epsilon_{\rm c},\,\,\,\,\,\cr
t^{\frac{2(\delta-3p)}{\delta+6}}\,\epsilon_\gamma^{-\frac{p}{2}},\hspace{0.7cm}\,\epsilon_{\rm c}<\epsilon_\gamma\,. \cr
}
\end{eqnarray}
}
Considering the particular value of $\delta=0$, the observable quantities derived in \cite{1999ApJ...519L..17S}  are recovered.
\paragraph{SSC Light curves.} Using the electron Lorentz factors (eq. \ref{eLor_syn_ism_l}) and the characteristic and cooling spectral breaks of the synchrotron emission  (eq. \ref{En_br_syn_ism_l}), the characteristic and cooling spectral breaks for SSC  in the fully adiabatic regime are
{\small
\bary\label{energies_break_ssc}
\epsilon^{\rm ssc}_{\rm m}&\simeq& 6.8\times10^{-8}\,{\rm eV}\,\, \left(\frac{1+z}{1.022}\right)^{\frac{12- \delta}{\delta+6}}\, g(p)^2\,\epsilon^4_{e,-2}\,\epsilon_{B,-4}^{\frac12}\,n_{-4}^{\frac{\delta-6}{2(\delta+6)}}\,\cr
&&\hspace{4.4cm} \times E_{52}^{\frac{6}{\delta+6}}\,  t_{10\,{\rm d}}^{-\frac{18}{\delta+6}}\cr
\epsilon^{\rm ssc}_{\rm c}&\simeq&  6.6\times 10^{8}\,{\rm TeV}\,\, \left(\frac{1+z}{1.022}\right)^{\frac{3(\delta - 4)}{\delta+6}}\,(1+Y)^{-4}\, \epsilon_{B,-4}^{-\frac72}\,\cr
&&\hspace{2.5cm} \times\,n_{-4}^{-\frac{7\delta+22}{2(\delta+6)}}\,E_{52}^{-\frac{10}{\delta+6}}\,  t_{10\,{\rm d}}^{-\frac{2(3-2\delta)}{\delta+6}}\,,
\eary
}
respectively.   Taking into account the maximum flux of the synchrotron radiation  (eq. \ref{flux_later}) and the optical depth \citep[see,][]{2001ApJ...548..787S}, the maximum flux emitted by the SSC process can be written as
{\small
\bary\label{flux_ssc_later}
F^{\rm ssc}_{\rm max} &\simeq& 2.1\times 10^{-22}\,{\rm mJy}\,\, \left(\frac{1+z}{1.022}\right)^{\frac{18-2\delta}{\delta+6}}\,\epsilon_{B,-4}^{\frac12}\,n_{-4}^{\frac{5(\delta+2)}{2(\delta+6)}}\,\cr
&&\hspace{3.4cm} \times  D^{-2}_{26.3} \,E_{52}^{\frac{10}{\delta+6}}\,  t_{10\,{\rm d}}^{\frac{2(2 \delta - 3)}{\delta+6}}\,.
\eary
}
Using the spectral breaks (eq. \ref{energies_break_ssc})  and the maximum flux (eq. \ref{flux_ssc_later}), then the light curves of the SSC process evolving  in the slow-cooling regime becomes
{\small
\begin{eqnarray}
\label{scsyn_t}
F^{\rm ssc}_{\nu}\propto\cases{
t^{\frac{4\delta}{\delta+6}}  \, \epsilon_\gamma^{\frac13},\hspace{1.8cm} \epsilon_\gamma<\epsilon^{\rm ssc}_{\rm m},\cr
t^{\frac{3 +4\delta-9p}{\delta+6}}\,\epsilon_\gamma^{-\frac{p-1}{2}},\hspace{0.7cm} \epsilon^{\rm ssc}_{\rm m}<\epsilon_\gamma<\epsilon^{\rm ssc}_{\rm c},\,\,\,\,\,\cr
t^{\frac{6+ 2\delta-9p}{\delta+6}}\,\epsilon_\gamma^{-\frac{p}{2}},\hspace{0.9cm} \,\epsilon^{\rm ssc}_{\rm c}<\epsilon_\gamma\,, \cr
}
\end{eqnarray}
}
respectively.\\
%
%
Figure \ref{fig1:jet_ism} shows the synchrotron and SSC light curves generated by the deceleration of the outflow in a uniform ISM-like medium. The left-hand panels show in solid lines the synchrotron fluxes  in radio at 1.4 GHz (magenta), optical at 1 eV (green) and  X-ray at 1 keV (gray) and the right-hand panels present the SSC fluxes in gamma-rays at 10 GeV (blue), $\gamma$-rays at 100 GeV (gold) for typical values of GRB afterglow parameters reported in the literature\footnote{The upper panels display the light curves for the values of $\tilde{E}=5\times 10^{52}\,{\rm erg}$, $n=10^{-4}\,{\rm cm^{-3}}$,  $\epsilon_B=10^{-2}$,  $\Delta \theta=20^\circ$ and $\alpha_s=2.1$ and the lower panels for the values of  $\tilde{E}=10^{51}\,{\rm erg}$,  $n=10^{-3}\,{\rm cm^{-3}}$, $\epsilon_B=10^{-4}$, $\Delta \theta=30^\circ$ and $\alpha_s=1.1$. In all panels  the values of  $\epsilon_e=0.1$, $p=2.2$ and $D=100\,{\rm Mpc}$ were used.}.   Dashed lines correspond to the sensitivities of the Expanded Very Large Array\footnote{https://public.nrao.edu/telescopes/vla/}  (EVLA) at 1.4 GHz (magenta),  the Large Synoptic Survey Telescope\footnote{https://www.lsst.org/} (LSST) at 1 eV (green),  X-ray Telescope on-board Swift satellite\footnote{https://swift.gsfc.nasa.gov/about\_swift/xrt\_desc.html}  (XRT) at 1 keV (gray),  Large Area Telescope on-board Fermi satellite \footnote{Data taken from \cite{2016CRPhy..17..617P}} (LAT) and MAGIC\footnote{Data taken from \cite{2008ApJ...687L...5T}}  at 100 GeV (gold).  The effect of the extragalactic background light (EBL) absorption proposed by \cite{2017A&A...603A..34F} is used to obtain the SSC light curves.With the parameter values used in these panels, the synchrotron and SSC light curves evolve in the slow-cooling regime. For another set of parameter values such as: a time scale of seconds, a higher uniform ISM-like  medium ($\geq 1 {\rm cm^{-3}}$) and equipartition parameters $\varepsilon_{\rm B}\sim 0.1$, the synchrotron and SSC light curves would evolve in the fast-cooling regime.  These panels shows that depending on the parameters used, the intensity of the fluxes will vary, but they will have similar behaviours in all electromagnetic bands;  they increase during the first $\sim$10- 50 days, then reach their respective maxima, and afterwards decrease.   It is worth mentioning that with the parameter values used in \cite{2019ApJ...871..123F} to model the electromagnetic counterpart of GW170817, the observed fluxes in X-ray, optical and radio bands increase during $\sim$ 120 days. The left-hand panels show that the evolution of synchrotron light curves  at  radio, optical  and X-ray bands  could or could not be detected by EVLA, LSST and Swift XRT, depending on the values of GRB afterglow parameters. For instance, the upper panel exhibits that these fluxes can be detected  whereas the lower panel displays the opposite case. On the other hand, the right-hand panels show that the evolution of SSC light curves at 10 and 100 GeV is too small to be detected by LAT and MAGIC observatories;  $\sim 10^4$ (upper) and $\sim 10^7$ (lower) times smaller. 
\subsection{Stratified stellar-wind like medium}
\begin{figure}[h!]
{ \centering
\resizebox*{1.\textwidth}{0.6\textheight}
{\includegraphics{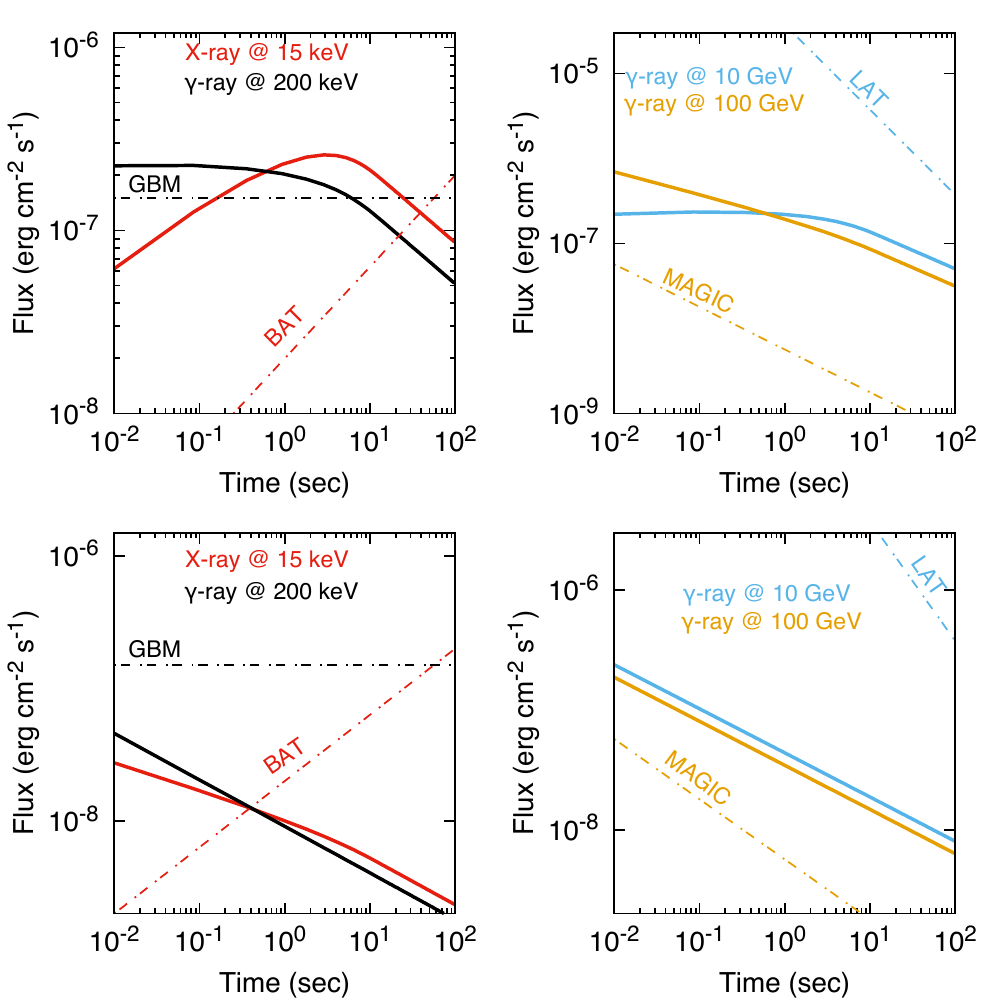}}
}
\caption{The figures show the light curves of the synchrotron (left-hand panels) and SSC (right-hand panels) fluxes radiated from the decelerated outflow in a wind-like density for the values of $E=10^{51}\,{\rm erg}$, $A_\star=10^4$,  $\epsilon_B=10^{-1}$, $\Delta \theta=15^\circ$ and $\alpha_s=1.1$ (upper panels) and  $\tilde{E}=10^{50}\,{\rm erg}$,  $A_\star=10^2\,$, $\epsilon_B=10^{-2}$,  $\Delta \theta=15^\circ$ and $\alpha_s=2.1$ (lower panels). The values of $\epsilon_e=0.1$, $p=2.2$ and $D=100 {\rm Mpc}$ were assumed in all the panels.}
 \label{fig2:jet_wind}
\end{figure}

 \cite{2014ApJ...784L..28N} numerically studied the jet propagation in the material ejected  by the neutrino-driven wind produced in the coalescence of a NS binary system. They used a density profile of the ejection along the pole given by {\small $\rho(r)\propto r^{-\lambda}$}.
Considering  the ejecta mass in the range of $10^{-3}\leq \frac{M_{\rm ej}}{ M_{\odot}} \leq 10^{-1}$ and onset times of jet injection up to $\sim 1$ s,  the authors found  the dynamics of the jet in the expanding ejecta with the power-law index of the density distribution $\lambda=3.5$.   \cite{2013ApJ...778L..16H} investigated the numerical results on the ejected material (its mass and total energy) for $\lambda$=2 and $\lambda$=3.  They found that the quantities depend weakly on the values of $\lambda$ and that  if the ejected mass increased by $\sim$ 10\% the value of $\lambda$ varies from $2\leq \lambda <3$. \cite{2013ApJ...773...78B} also investigated the dynamics of the ejected mass  of a merger of two NSs. They argued that the circumburst medium in the close vicinity of a merger could be approximated as a wind medium with a density given by the power-law $\rho \propto r^{-2}$.  \cite{2018ApJ...863L..34B} analyzed the GBM data of the short GRB 150101B and used $\lambda=2$ to explain the observed gamma-ray flux.\\
 In this work  the value of $\lambda=2$ will be chosen for our analysis.  Taking into account that the lateral expansion phase is expected to occur pretty far from the close vicinity of the merger, then the lateral expansion phase in a wind-like medium is not considered.  Therefore, we only derive the synchrotron and SSC light curves in the relativistic phase.    In the case of a stratified stellar-wind like medium, the number density is given by  $n(r)=\frac{\rho(r)}{m_p}=\frac{A}{m_p}\,r^{-2}$ where $A=\frac{\dot{M}}{4\pi\, v}=\,5\times 10^{11}\,A_\star\,{\rm g\,cm^{-1}}$, with $\dot{M}$  the mass-loss rate, $v$ the velocity of the outflow and $A_\star$ a density parameter \citep[e.g., see][]{2016ApJ...818..190F, 2017ApJ...837..116B}. \\
Taking into account the Blandford-McKee solution for a stratified stellar-wind like medium,  the bulk Lorentz  factor derived through the adiabatic evolution \citep{1976PhFl...19.1130B, 1997ApJ...489L..37S} is given by
{\small
\be\label{Gamma_wind}
\Gamma=  16.4 \left(\frac{1+z}{1.022}\right)^{\frac{1}{\delta+4}}\, \xi^{-\frac{2}{\delta+4}}\, A_{\star,-1}^{-\frac{1}{\delta+4}}\, \Delta \theta^{-\frac{6}{\delta+4}}_{15^\circ}  \,  \tilde{E}^{\frac{1}{\delta+4}}_{52}  \,t^{-\frac{1}{\delta+4}}_{1\,\rm s}\,,
\ee
}
with the fiducial energy given by {\small $\tilde{E}=  \frac{16\pi}{3}\, (1+z)^{-1}\,\xi^2\, A_\star\, \Delta \theta^6\, \Gamma^{\delta+4}\,t\,$}.
\paragraph{Synchrotron Light curves.}  Using the bulk Lorentz factor (eq. \ref{Gamma_wind})  and the synchrotron afterglow theory for a wind-like medium \citep{2000ApJ...536..195C, 2000ApJ...543...66P}, we derive the relevant quantities of synchrotron emission for our model  in the fully adiabatic regime.  The minimum and cooling electron Lorentz factors are  given by
{\small
\bary\label{eLor_syn_wind}
\gamma_m&=& 42.1\left(\frac{1+z}{1.022}\right)^{\frac{1}{\delta+4}}\, \xi^{-\frac{2}{\delta+4}}\,g(p)\, \epsilon_{\rm e,-2}\,\Delta\theta^{\frac{-6}{\delta+4}}_{15^\circ} \, A_{\star,-1}^{-\frac{1}{\delta+4}}\, \tilde{E}^{\frac{1}{\delta+4}}_{52}\,\cr 
&&\hspace{6.3cm} \times \,t^{-\frac{1}{\delta+4}}_{\rm 1s} \cr
\gamma_c&=& 45.6 \left(\frac{1+z}{1.022}\right)^{-\frac{\delta+3}{\delta+4}} (1+Y)^{-1}\,\xi^{\frac{2(\delta+3)}{\delta+4}}\, \epsilon_{B,-4}^{-1}\,A_{\star,-1}^{-\frac{\delta+5}{\delta+4}}\,\cr
&&\hspace{3.5cm} \times \Delta\theta^{-\frac{6}{\delta+4}}_{15^\circ} \,\tilde{E}^{\frac{1}{\delta+4}}_{52}\,t^{\frac{\delta+3}{\delta+4}}_{\rm 1s}\,, 
\eary
}
which correspond to a comoving magnetic field given by
{\small
\bary
B'&\simeq&9.5\times 10^2\,\,{\rm mG}\,\left(\frac{1+z}{1.022}\right)^{-\frac{\delta+5}{\delta+4}}\,\xi^{-\frac{2(\delta+3)}{\delta+4}} \epsilon^{\frac12}_{B,-4}\, A_{\star,-1}^{\frac{\delta+6}{2(\delta+4)}}\, \Delta\theta^{\frac{6}{\delta+4}}_{15^\circ}\cr
&&\hspace{5cm} \times \,E^{-\frac{1}{\delta+4}}_{52}\,t_{1\,{\rm d}}^{-\frac{\delta+3}{\delta+4}}\,.
\eary
}%
The characteristic and cooling spectral breaks for synchrotron emission are
{\small
\bary\label{En_br_syn_wind}
\epsilon^{\rm syn}_{\rm m}&\simeq& 0.2\,{\rm eV}\,\, \left(\frac{1+z}{1.022}\right)^{\frac{2}{\delta+4}}\, \xi^{-\frac{2(\delta+6)}{\delta+4}}  \,\epsilon^2_{\rm e,-2}\,\epsilon_{B,-4}^{\frac12}\cr
&&\hspace{2cm} \times\,\,A_{\star,-1}^{\frac{\delta}{2(\delta+4)}}\, \,\Delta\theta_{15^\circ}^{-\frac{12}{\delta+4}}\, E_{52}^{\frac{2}{\delta+4}}\,  t_{1\,{\rm s}}^{-\frac{\delta+6}{\delta+4}}\cr
\epsilon^{\rm syn}_{\rm c}&\simeq&  0.1\,{\rm eV}\,\, \left(\frac{1+z}{1.022}\right)^{-\frac{2(\delta +3)}{\delta+4}}\,\xi^{\frac{2(\delta+2)}{\delta+4}}\,(1+Y)^{-2}\,\cr
&&\hspace{1.3cm} \times \,\epsilon_{B,-4}^{-\frac32} A_{\star,-1}^{-\frac{3\delta+16}{2(\delta+8)}}\,\Delta\theta_{15^\circ}^{-\frac{12}{\delta+4}}\,E_{52}^{\frac{2}{\delta+4}}\,  t_{1\,{\rm s}}^{\frac{\delta+2}{\delta+4}}\,,
%
\eary
}
respectively.   Given  the maximum emissivity  in a stratified stellar-wind like medium,  the maximum flux radiated by synchrotron emission is given by
{\small
\bary\label{Flux_syn_wind}
F^{\rm syn}_{\rm max} &\simeq& 1.9\times 10^{3}\,{\rm mJy}\,\, \left(\frac{1+z}{1.022}\right)^{\frac{2(\delta+5)}{\delta+4}}\,  \xi^{-\frac{4}{\delta+4}} \,\epsilon_{B,-4}^{\frac12}\,A_{\star,-1}^{\frac{3\delta+8}{2(\delta+4)}}\,\cr
&&\hspace{2cm} \times D^2_{26.3}\,\Delta\theta_{15^\circ}^{-\frac{12}{\delta+4}}\,E_{52}^{\frac{2}{\delta+4}}\,  t_{1\,{\rm s}}^{-\frac{2}{\delta+4}}\,.
\eary
}
Using the synchrotron spectral breaks (eq. \ref{En_br_syn_wind}) and the maximum flux  (eq. \ref{Flux_syn_wind}), the synchrotron light curves in the fast- and slow-cooling regime can be written  as
{\small
\begin{eqnarray}\label{Fl_syn_wind_fc}
F^{\rm syn}_{\nu}\propto\cases{
t^{-\frac{\delta+8}{3(\delta+4)}}  \, \epsilon_\gamma^{\frac13},\hspace{1.3cm} \epsilon_\gamma<\epsilon^{\rm syn}_{\rm c},\cr
t^{\frac{\delta-2}{2(\delta+4)}}\,\epsilon_\gamma^{-\frac{p-1}{2}},\hspace{1.0cm}\epsilon^{\rm syn}_{\rm c}<\epsilon_\gamma<\epsilon^{\rm syn}_{\rm m},\,\,\,\,\,\cr
t^{\frac{4+2\delta-p\delta-6p}{2(\delta+4)}}\,\epsilon_\gamma^{-\frac{p}{2}},\hspace{0.6cm}\epsilon^{\rm syn}_{\rm m}<\epsilon_\gamma\,, \cr
}
\end{eqnarray}
}
and
{\small
\begin{eqnarray}\label{Fl_syn_wind_sc}
F^{\rm syn}_{\nu}\propto\cases{
t^{\frac{\delta}{3(\delta+4)}}  \, \epsilon_\gamma^{\frac13},\hspace{1.6cm} \epsilon_\gamma<\epsilon^{\rm syn}_{\rm m},\cr
t^{\frac{2+\delta-6p-p\delta}{2(\delta+4)}}\,\epsilon_\gamma^{-\frac{p-1}{2}},\hspace{0.5cm} \epsilon^{\rm syn}_{\rm m}<\epsilon_\gamma<\epsilon^{\rm syn}_{\rm c},\,\,\,\,\,\cr
t^{\frac{4+2\delta-p\delta-6p}{2(\delta+4)}}\,\epsilon_\gamma^{-\frac{p}{2}},\hspace{0.7cm} \epsilon^{\rm syn}_{\rm c}<\epsilon_\gamma\,, \cr
}
\end{eqnarray}
}
respectively.   Considering the particular scenario of $\delta=0$, the observable quantities derived in  \citep{2000ApJ...536..195C, 2000ApJ...543...66P} and the light curves explicitly shown in \cite{2015ApJ...804..105F} are recovered.\
\paragraph{SSC Light curves.} Using eqs. (\ref{eLor_syn_wind}) and (\ref{En_br_syn_wind}), the characteristic and cooling spectral breaks  of SSC emission are
{\small
\bary\label{energies_break}
\epsilon^{\rm ssc}_{\rm m}&\simeq& 0.1\times 10^{-3}\,{\rm eV}\,\, \left(\frac{1+z}{1.022}\right)^{\frac{4}{\delta+4}}\, \xi^{-\frac{2(\delta+8)}{\delta+4}}    \,\epsilon^4_{\rm e,-2}\,\epsilon_{B,-4}^{\frac12}\,\cr
&&\hspace{1.6cm} \times\,A_{\star,-1}^{\frac{\delta-4}{2(\delta+4)}}\,\Delta\theta_{15^\circ}^{-\frac{24}{\delta+4}}\,E_{52}^{\frac{4}{\delta+4}}\,  t_{1\,{\rm s}}^{-\frac{\delta+8}{\delta+4}}\cr
\epsilon^{\rm ssc}_{\rm c}&\simeq&  0.2\,{\rm keV}\,\, \left(\frac{1+z}{1.022}\right)^{-\frac{4(\delta +3)}{\delta+4}}\,\xi^{\frac{2(3\delta+8)}{\delta+4}}\,(1+Y)^{-4}\cr
&&\hspace{1.2cm} \times\,\epsilon_{B,-4}^{-\frac72}\,A_{\star,-1}^{-\frac{7\delta+36}{2(\delta+4)}}\,\Delta\theta_{15^\circ}^{-\frac{24}{\delta+4}}\,E_{52}^{\frac{4}{\delta+4}}\,  t_{1\,{\rm s}}^{\frac{3\delta+8}{\delta+4}}\,, 
%
\eary
}
respectively.  The break energy due to  the Klein-Nishina (KN) effect  is given by
{\small
\bary
\epsilon^{\rm ssc}_{\rm KN}&\simeq& 0.3\,{\rm GeV}\,\, \left(\frac{1+z}{1.022}\right)^{-\frac{2(\delta + 3)}{\delta+4}}\,(1+Y)^{-1}\, \xi^{\frac{2(\delta+2)}{\delta+4}}\, \epsilon_{B,-4}^{-1}\,\cr
&&\hspace{2.2cm} \times\,A_{\star,-1}^{-\frac{\delta+6}{\delta+4)}}\,\Delta\theta_{15^\circ}^{-\frac{12}{\delta+4}}\,E_{52}^{\frac{2}{\delta+4}}\,  t_{1\,{\rm s}}^{\frac{\delta+2}{\delta+4}}\,.
\eary
}
Considering the maximum flux of synchrotron emission (eq. \ref{Flux_syn_wind}), the maximum flux emitted by the SSC process can be written as
{\small
\bary \label{Flux_ssc_wind}
F^{\rm ssc}_{\rm max} &\simeq& 19.2\,{\rm mJy}\,\left(\frac{1+z}{1.022}\right)^3\,  \xi^{-2} \,\epsilon_{B,-4}^{\frac12}\,A_{\star,-1}^{\frac52}\,D^{-2}_{26.3}\,  t_{1\,{\rm s}}^{-1}.
\eary
}
Using the characteristic and cooling spectral breaks (eq. \ref{energies_break}) and the maximum flux (eq. \ref{Flux_ssc_wind}),  the light curves in the fast- and slow-cooling regime are
{\small
\begin{eqnarray}
\label{Fl_ssc_wind_fc}
F^{\rm ssc}_{\nu}\propto\cases{
t^{-\frac{2(3\delta+10)}{3(\delta+4)}}  \, \epsilon_\gamma^{\frac13},\hspace{1.3cm} \epsilon_\gamma<\epsilon^{\rm ssc}_{\rm c},\cr
t^{\frac{\delta}{2(\delta+4)}}\,\epsilon_\gamma^{-\frac{1}{2}},\hspace{1.5cm} \epsilon^{\rm ssc}_{\rm c}<\epsilon_\gamma<\epsilon^{\rm ssc}_{\rm m},\,\,\,\,\,\cr
t^{\frac{8+2\delta-8p-p\delta}{2(\delta+4)}}\,\epsilon_\gamma^{-\frac{p}{2}},\hspace{0.8cm} \epsilon^{\rm ssc}_{\rm m}<\epsilon_\gamma\,, \cr
}
\end{eqnarray}
}
and
{\small
\begin{eqnarray}
\label{Fl_ssc_wind_sc}
F^{\rm ssc}_{\nu}\propto \cases{
t^{-\frac{2(\delta+2)}{3(\delta+4)}}  \, \epsilon_\gamma^{\frac13},\hspace{1.3cm} \epsilon_\gamma<\epsilon^{\rm ssc}_{\rm m},\cr
t^{-\frac{\delta +8p + p\delta}{2(\delta+4)}}\,\epsilon_\gamma^{-\frac{p-1}{2}},\hspace{0.5cm} \epsilon^{\rm ssc}_{\rm m}<\epsilon_\gamma<\epsilon^{\rm ssc}_{\rm c},\,\,\,\,\,\cr
t^{\frac{8+2\delta-8p -p\delta}{2(\delta+4)}}\,\epsilon_\gamma^{-\frac{p}{2}},\hspace{0.6cm} \epsilon^{\rm ssc}_{\rm c}<\epsilon_\gamma\,, \cr
}
\end{eqnarray}
}
respectively.\\
Figure \ref{fig2:jet_wind} shows the light curves of the synchrotron (left-hand panels) and SSC (right-hand panels) fluxes radiated from the decelerated off-axis jet for typical parameter values of a GRB evolving in a stratified stellar-wind like medium\footnote{The upper panels show the light curves for values of  $\tilde{E}=10^{51}\,{\rm erg}$, $A_\star=10^4$,  $\epsilon_B=10^{-1}$, $\Delta \theta=15^\circ$ and $\alpha_s=1.1$, and the lower panels for the values of $\tilde{E}=10^{50}\,{\rm erg}$,  $A_\star=10^2\,$, $\epsilon_B=10^{-2}$,  $\Delta \theta=15^\circ$ and $\alpha_s=2.1$.  In all panels the values of $\epsilon_e=0.1$, $p=2.2$ and $D=100 {\rm Mpc}$ were used.}.   The synchrotron and SSC light curves are displayed for two electromagnetic bands and for the chosen parameters these correspond to earlier times than Figure \ref{fig1:jet_ism}. For synchrotron emission: X-rays at  15 keV and  $\gamma$-rays at 200 keV and for SSC emission:  $\gamma$-rays at 10 and 100 GeV.      Dashed lines correspond to the sensitivities of the GBM on-board the Fermi satellite at 200 keV (black) and Burst Alert Telescope (BAT) on-board Swift satellite at 15 keV (red)\footnote{https://swift.gsfc.nasa.gov/proposals/tech\_appd/swiftta\_v12/node25.html},  Fermi LAT\footnote{Data taken from \cite{2016CRPhy..17..617P}} and MAGIC\footnote{Data taken from \cite{2008ApJ...687L...5T}}  at 100 GeV (gold).  The effect of the EBL absorption introduced in \cite{2017A&A...603A..34F} is used to obtain the SSC light curves. With the parameter values used in these panels, the synchrotron and SSC light curves evolve in the fast-cooling regime. For another set of parameter values such as: a time scale of hours and equipartition parameters $\varepsilon_{\rm B}\sim 0.4$, the synchrotron and SSC light curves would evolve in the slow-cooling regime.    These panels shows that depending on the instrument used to detect the electromagnetic emission and the parameters introduced in the model, the observed fluxes will have distinct behaviors. For instance, the synchrotron flux observed at 15 keV  increases during the first $\sim$ 5 seconds, then reaches its respective maximum, and decreases afterwards, and  the synchrotron flux observed at 200 keV is almost constant during the first second and then starts to decrease. The SSC flux observed at 10 GeV remains constant during the first 3 seconds and then decreases, and at 100 GeV, it decreases monotonically.      The upper panel shows that the evolution of synchrotron light curves  at  X-ray and $\gamma$-ray bands  could be detected during the first $\sim$ 5 - 10 s and the lower panel shows that Swift BAT could detect the synchrotron emission up to 0.2 s  and Fermi BAT could not have detected the $\gamma$-ray emission.  The right-hand panels show that SSC emission cannot be observed by Fermi LAT whereas it can be detected by the MAGIC telescope irrespective of the parameter values used.  
\section{Applications}
\subsection{GRB 080503}
\begin{figure}
{ \centering
\resizebox*{0.7\textwidth}{0.4\textheight}
{\includegraphics{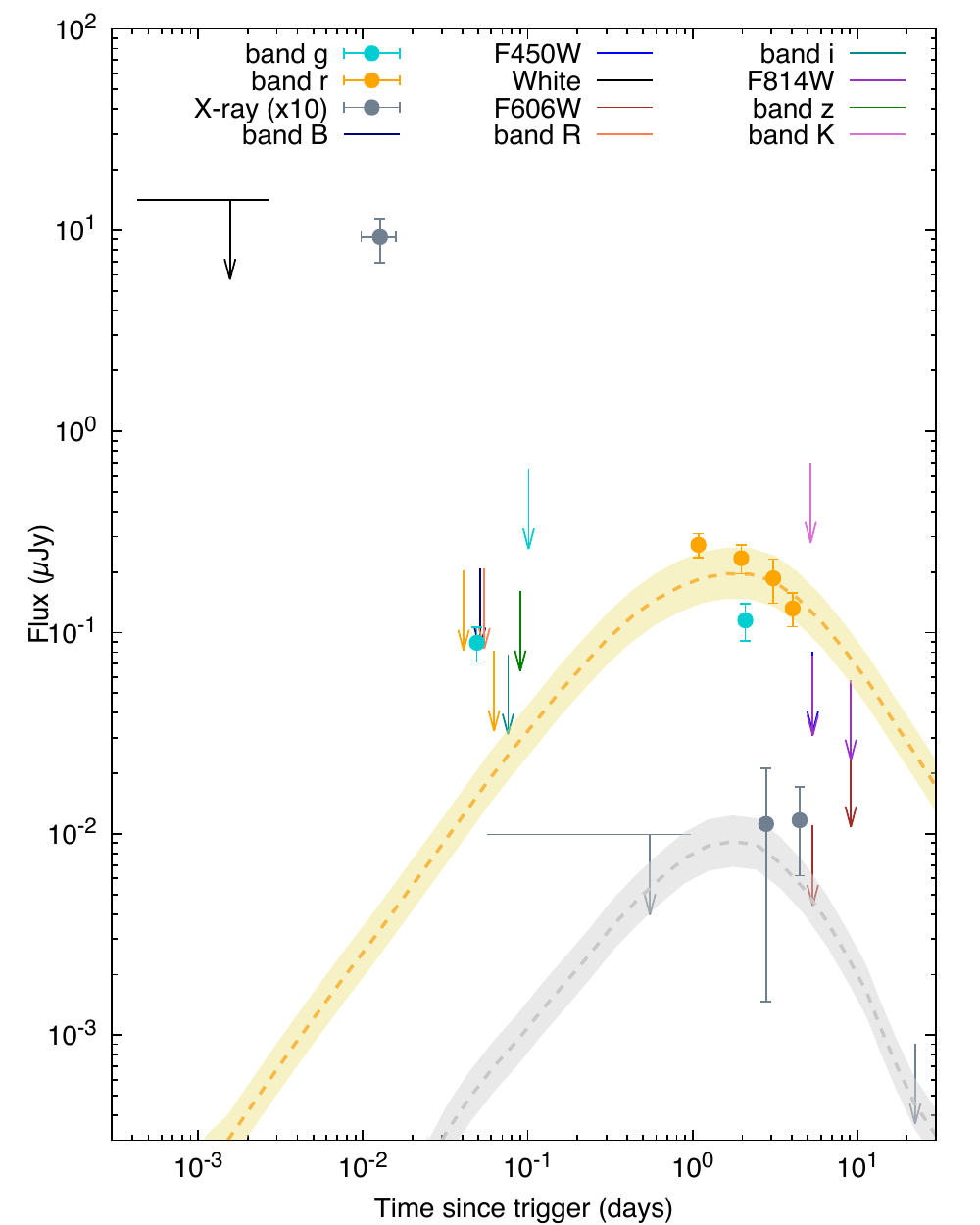}}
}
\caption{The best-fit synchrotron light curves generated when the outflow is decelerated in a uniform  ISM-like medium. These synchrotron light curves are displayed at the optical (yellow) and X-ray (gray) energy bands with the data points and upper limits of GRB 080503. Data are taken from \cite{2009ApJ...696.1871P}.  The best-fit values of the parameters used in our model for optical  (column 2) and X-ray (column 3) bands are reported in Table \ref{parameters}.}
\label{LC_GRB080503}
\end{figure}

GRB 080503 triggered Swift BAT at 2008 May 3 12:26:13 UTC.   The duration and the observed flux of the initial main spike in the energy range of 15 - 150 keV were $0.32\pm 0.07\,{\rm s}$ and $(1.2\pm 0.2)\times 10^{-7}\,{\rm erg\,cm^{-2}\,s^{-1}}$, respectively.   Details of the X-ray and optical afterglow observations collected with Swift,  Chandra,  Keck-I, Gemini-N and  Hubble Space Telescope (HST) are reported in \cite{2009ApJ...696.1871P}.\\
\\
To obtain the best-fit values of the parameters that describe the optical and X-ray data with their upper limits of GRB 080503,  we use the Bayesian statistical method of Markov-Chain Monte Carlo (MCMC) simulations \citep[e.g see,][]{2019arXiv190406976F}.   The model can be explained by a set of eight parameters, \{$\tilde{E}$, $n$,  $p$,  $\Delta\theta$,  $\varepsilon_e$, $\varepsilon_{B}$, $k$ and $\alpha_s$\}.   To describe the whole data, a total of 16500 samples and 4200 tuning steps were run.  All parameters are described by normal distributions.  The best-fit values and the median of the posterior distributions of the parameters are exhibited in corner plots, as shown in Figures  \ref{GRB080305_optical} and \ref{GRB080305_X-ray} for optical and X-ray data, respectively.   The best-fit values in these figures are shown in green color and the median of the posterior distributions are  reported in Table \ref{parameters}.\\ 
Figure \ref{LC_GRB080503} shows the optical and X-ray light curves with the fits and uncertainties obtained with the synchrotron forward-shock model evolving in a homogeneous density.  The non-thermal optical and X-ray observations  are consistent with the outflow described by eq. \ref{Eoff}.  It would suggest that multiwavelength observations were generated at the same emitting region and by the same radiative process.   The  peak time in the observed flux at  $\sim$ one day  and after the fast decay is consistent with the fact that the  beaming cone of the synchrotron radiation reaches our line of sight. The best-fit values of the parameters for optical (column 2) and X-ray (column 3) are reported in Table \ref{parameters}.\\
The value of the homogeneous medium required to describe the non-thermal long-lasting emission indicates that the progenitor of GRB 080503, like other sGRBs, exploded in a very low density environment.  The very low density is in agreement  with the larger offsets of sGRBs compared with long GRBs.\\  
The value of the spectral index of the electron population is consistent with the typical value reported when relativistic electrons accelerated in the forward shocks are cooled down by synchrotron radiation \citep[see, e.g.][]{2015PhR...561....1K}. It reaffirms that this emission was originated in the GRB afterglow.\\
Although significant efforts to find the jet breaks in sGRBs have been made, only few detections have been successful.  Given these detections, \cite{2014ARA&A..52...43B} showed that the mean of the jet breaks lies around $\theta_j\approx 3^\circ - 6^\circ$.  Assuming a value of $4^\circ$ for GRB 080503, the viewing angle becomes $\theta_{\rm obs}\approx 3^\circ$.   Given the observed fluxes of the main spike reported by Swift BAT \citep{2009ApJ...696.1871P} during the first second and  the long-lasting emission with a timescale of days, it can be seen that the main spike is fainter than the long-lasting afterglow emission.  We argue that the main spike component was viewed nearly off-axis whereas the component associated to the long-lasting afterglow emission was viewed more widely beamed. \\
\cite{2009ApJ...696.1871P} analyzed the optical and the X-ray observations at $\sim$ one day. Pointing out that the X-ray and optical observations exhibited similar evolutions, authors discarded the kilonova-like emission proposed by \cite{1998ApJ...507L..59L} and provided an afterglow interpretation.  They proposed that the faint afterglow relative to the bright prompt emission could be explained in term of the very low circumburst medium and also argued that the late optical and X-ray bumps could be interpreted in the framework of a slightly off-axis jet or a refreshed shock.  \cite{2012A&A...541A..88H} showed that the origin of the late rebrightening in GRB 080503 could be due to refreshed shocks.  \cite{2015ApJ...807..163G} argued that the late optical and X-ray bump was consistent with the emission from a magnetar-powered ``merger-nova".  Our analysis indicates  that the X-ray and optical observations  at $\sim$ one day are consistent with the afterglow emission seen slightly off-axis.

\subsection{GRB 140903A}
\begin{figure}
{ \centering
\resizebox*{0.7\textwidth}{0.4\textheight}
{\includegraphics{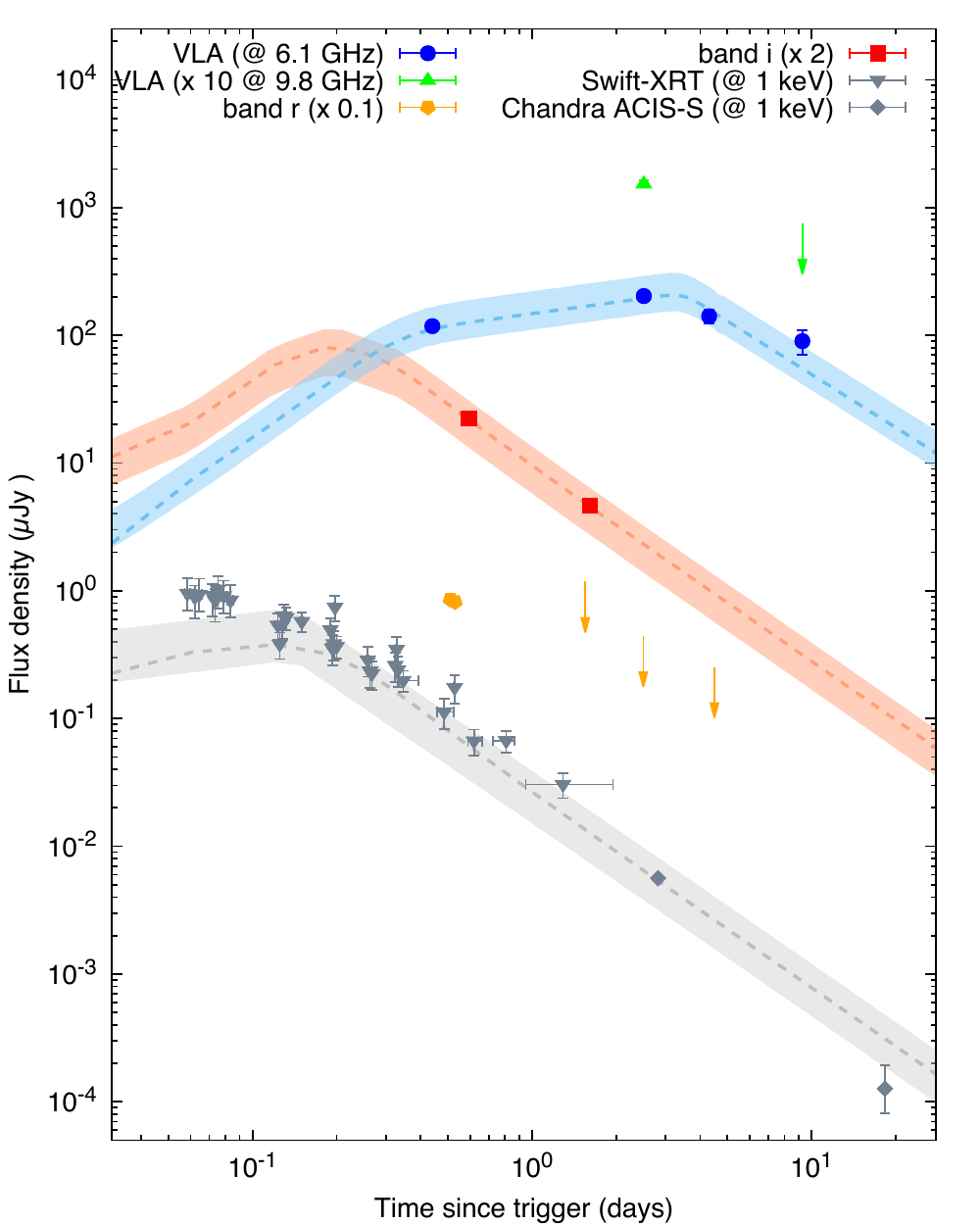}}
}
\caption{The best-fit synchrotron light curves generated when the outflow is decelerated in a uniform ISM-like medium. These synchrotron light curves are displayed at radio (blue), optical (red) and X-ray (gray) energy bands with the data points and upper limits of GRB 140903A. Data are taken from \cite{2016ApJ...827..102T}.  The best-fit values of the parameters used in our model for radio (column 4), optical (column 5) and X-ray (column 6) bands are reported in Table \ref{parameters}.}
\label{LC_GRB140903A}
\end{figure}

GRB 140903A  was detected by  the Swift BAT at 15:00:30 UT on 2014 September 14.  Details of the X-ray, optical and radio afterglow observations collected with Swift,  Chandra,  different optical telescopes and Very Large Area (VLA) are reported in \cite{2016ApJ...827..102T}.\\
To obtain the best-fit values of the parameters that adjust the radio, optical and X-ray observations of GRB 140903A,  once again we used the MCMC simulations.  In this case,  a total of 16600 samples and 4300 tuning steps were performed to describe the whole data.  The best-fit values and the median of the posterior distributions of the parameters are exhibited in Figures \ref{GRB140903A_radio}, \ref{GRB140903A_optical} and \ref{GRB140903A_X-ray}, respectively.  The best-fit values are shown in green color and the median of the posterior distributions are  reported in Table \ref{parameters}. The best-fit values of  radio, optical and X-ray data  are shown in columns 4, 5 and 6, respectively.\\
Figure \ref{LC_GRB140903A} shows the radio, optical and X-ray light curves of GRB 140903A with the fits  obtained with the synchrotron forward-shock model evolving in a homogeneous density.  Taking into account a typical value of $2^\circ - 4^\circ$  \citep{2014ARA&A..52...43B}, the viewing angle becomes $\theta_{\rm obs}\approx 3^\circ$.   These values suggest that the jet is seen slightly off-axis.  The viewing angle and  the best-fit values of the spectral index of the electron population $p=2.4$, the microphysical parameters $\epsilon_{\rm e}\simeq 9\times10^{-2}$ and $\epsilon_{\rm B}\simeq 8\times10^{-2}$ are similar to  the ones reported in \cite{2016ApJ...827..102T}.   The value of the power-law index of the electron population  indicates that the long-lasting emission was originated in the GRB afterglow. The derived value  of the kinetic energy $\sim 10^{51}\,{\rm erg}$  suggests that pair annihilation of $\nu$s and $\bar{\nu}$s  is a possible mechanism to provide the energy budget $L_{\nu\bar\nu}\lesssim 10^{51}\, {\rm erg s^{-1}}$. This result agrees with numerical simulation of merging NS-NS or NS-BH systems \citep{2004MNRAS.352..753S, 2007A&A...463...51B, 2013ApJ...762L..18G, 2013ApJ...771L..26G}.\\  
\cite{2016ApJ...827..102T} reported and gave a complete analysis of the afterglow observations up to $\sim$ 15 days of GRB 140903A.  Requiring the fireball scenario, authors showed that this burst was originated from a collimated jet seen off-axis  and also  associated to a compact binary object. \cite{2017ApJ...835...73Z} attributed the X-ray ``plateau" exhibited in GRB140903A to the energy injection into the decelerating blast wave and then authors modelled the late afterglow emission requiring a jet opening angle of $\approx 3^\circ$.  Our analysis leads to GRB 140903A  was generated in a collimated jet seen off-axis that decelerates in a homogeneous density.
\vspace{1cm}
\subsection{GRB 150101B }
\begin{figure}[h!]
{\centering
\resizebox*{0.5\textwidth}{0.35\textheight}
{\includegraphics{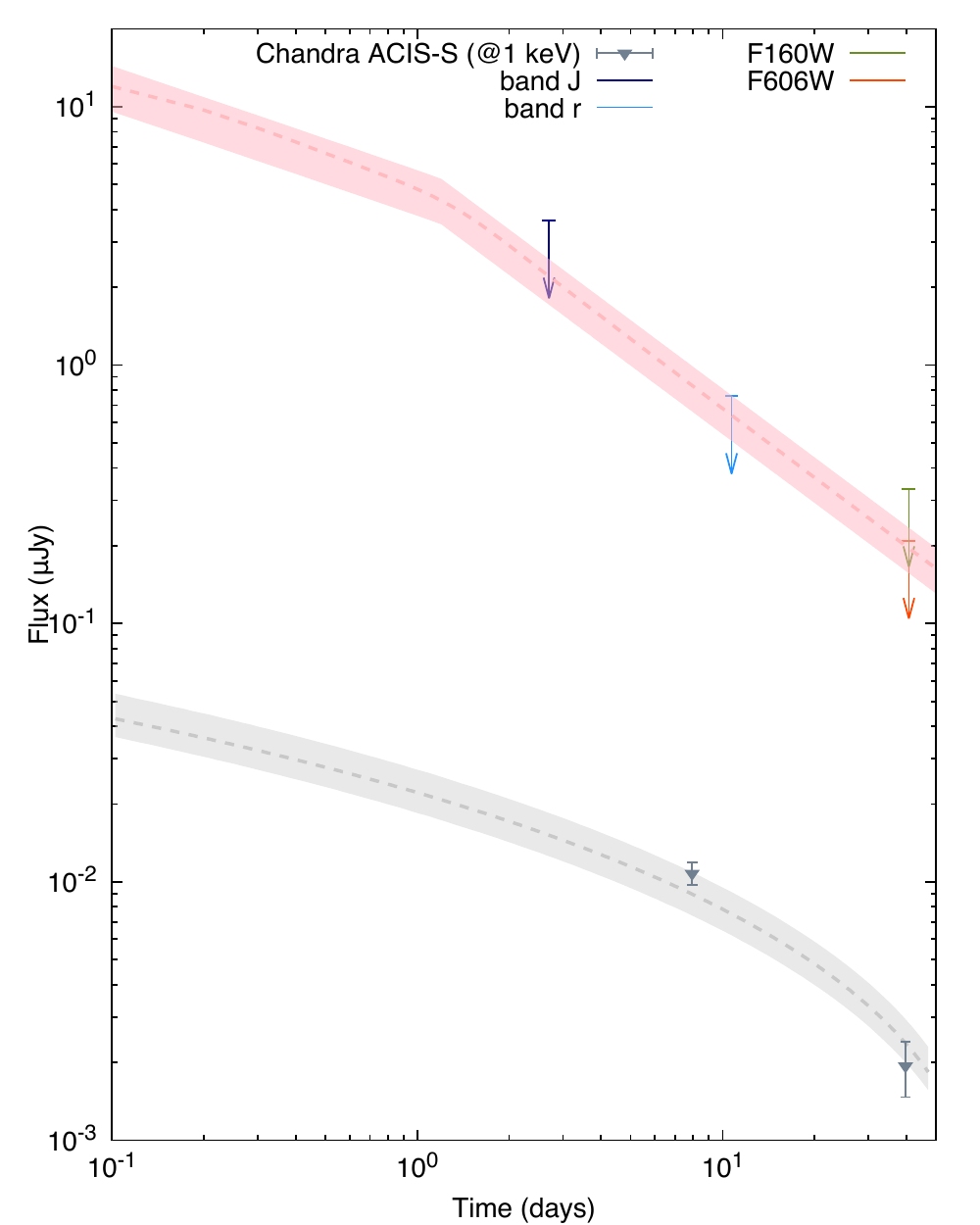}}
\resizebox*{0.5\textwidth}{0.35\textheight}
{\includegraphics{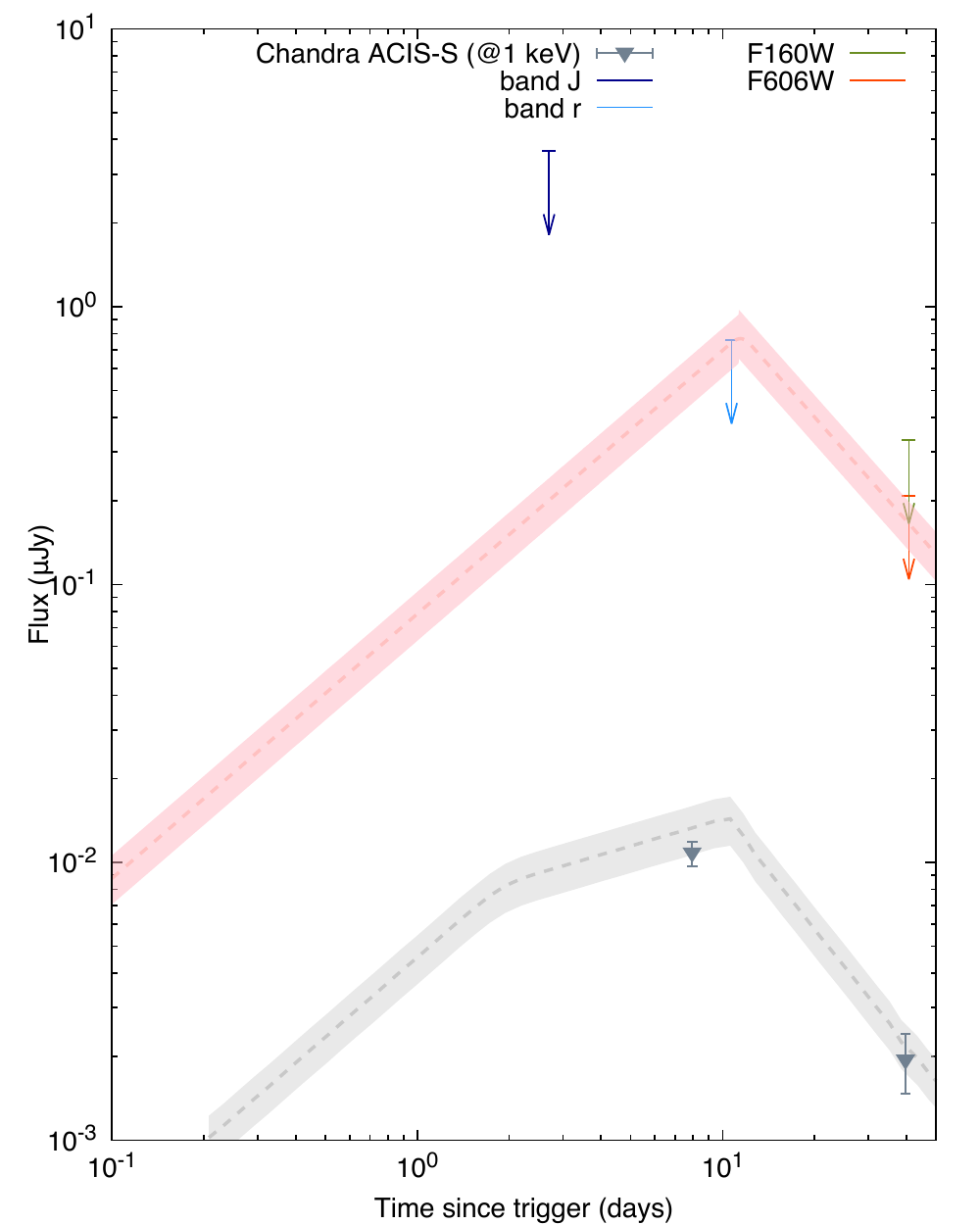}}
}
\caption{The best-fit synchrotron light curves generated when the outflow is decelerated in a wind- (right) and a uniform ISM-like (left) medium. These synchrotron light curves are displayed at optical (red) and X-ray (gray) energy bands with the data points and upper limits of GRB 150101B.  X-ray data is taken from  \cite{2016ApJ...833..151F} and optical upper limits are taken from \cite{2018NatCo...9.4089T}.  The best-fit values of the parameters used in our model for X-rays (columns 7 and 8) are reported in Table \ref{parameters}.}
\label{LC_GRB150101B}
\end{figure}

%
The Swift BAT and Fermi GBM detected GRB 150101B at 15:23:35  and 15:24:34.468 UT on 2015 January 01, respectively \citep{2018ApJ...863L..34B}.   Data analysis of  Swift BAT  revealed a bright  $\gamma$-ray pulse with duration and fluence of $T_{90}=0.012\pm0.001$ s and $F_{\gamma}=(6.1\pm 2.2)\times 10^{-8}\,{\rm erg\, cm^{-2}}$, respectively.  Details of the X-ray and optical afterglow observations with their upper limits are reported in  \cite{2016ApJ...833..151F} and \cite{2018NatCo...9.4089T}.\\
To obtain the best-fit values of the parameters that adjust the X-ray and optical observations of GRB 150101B,  once again we use the MCMC simulations.  In this case,  a total of 16400 samples and 4300 tuning steps were performed to describe the whole data.  The best-fit values and the median of the posterior distributions of the parameters are exhibited in Figure \ref{GRB150101B_X-ray}.  The best-fit values are shown in green color and the median of the posterior distributions are  reported in Table \ref{parameters} (columns  7 and 8).\\ 
\\
Figure \ref{LC_GRB150101B} shows the X-ray light curve of GRB 150101B with the fit and uncertainties obtained with the synchrotron forward-shock model evolving in a wind (left) and homogeneous (right) density.  As the homogeneous density is considered, the values of the spectral index of the electron population, the circumburst density, the microphysical parameters and the viewing angle disfavor the isotropic cocoon model reported in  \cite{2018NatCo...9.4089T} and are consistent with the values of a structured jet.  \cite{2016ApJ...833..151F} modeled the evolution of the afterglow observations in GRB 150101B and estimated the isotropic-equivalent kinetic energy of $\approx (6-14)\times 10^{51}\,{\rm erg}$ and a jet opening angle of $\gtrsim 9^\circ$.   Our analysis leads to similar values of kinetic energy and a jet opening angle.   Given the observed flux of the short and hard spike reported by Fermi GBM  \citep{2018ApJ...863L..34B} and  the X-ray afterglow emission detected in a timescale of days,  it can be observed that the short and hard spike is fainter than the X-ray emission.  We conclude that the bright spike component was viewed nearly off-axis whereas the long-lasting emission was viewed more widely beamed.  The best-fit value  of the circumburst medium obtained  suggests that the progenitor of GRB 150101B, like other short bursts, exploded in a low density environment.  When the wind-like medium is considered, our model can describe consistently the X-ray data and optical upper limits. In this case, the value of the equivalent kinetic energy is less and the magnetic microphysical parameter is larger than those derived assuming a homogeneous medium.  The result of the density parameter derived with our model  is consistent with the allowed range of values reported by the GBM collaboration \citep{2018ApJ...863L..34B} after describing the short and hard gamma-ray peaks.
\vspace{0.5cm}
\subsection{GRB 160821B}
\begin{figure}[h!]
{\centering
\resizebox*{0.5\textwidth}{0.35\textheight}
{\includegraphics{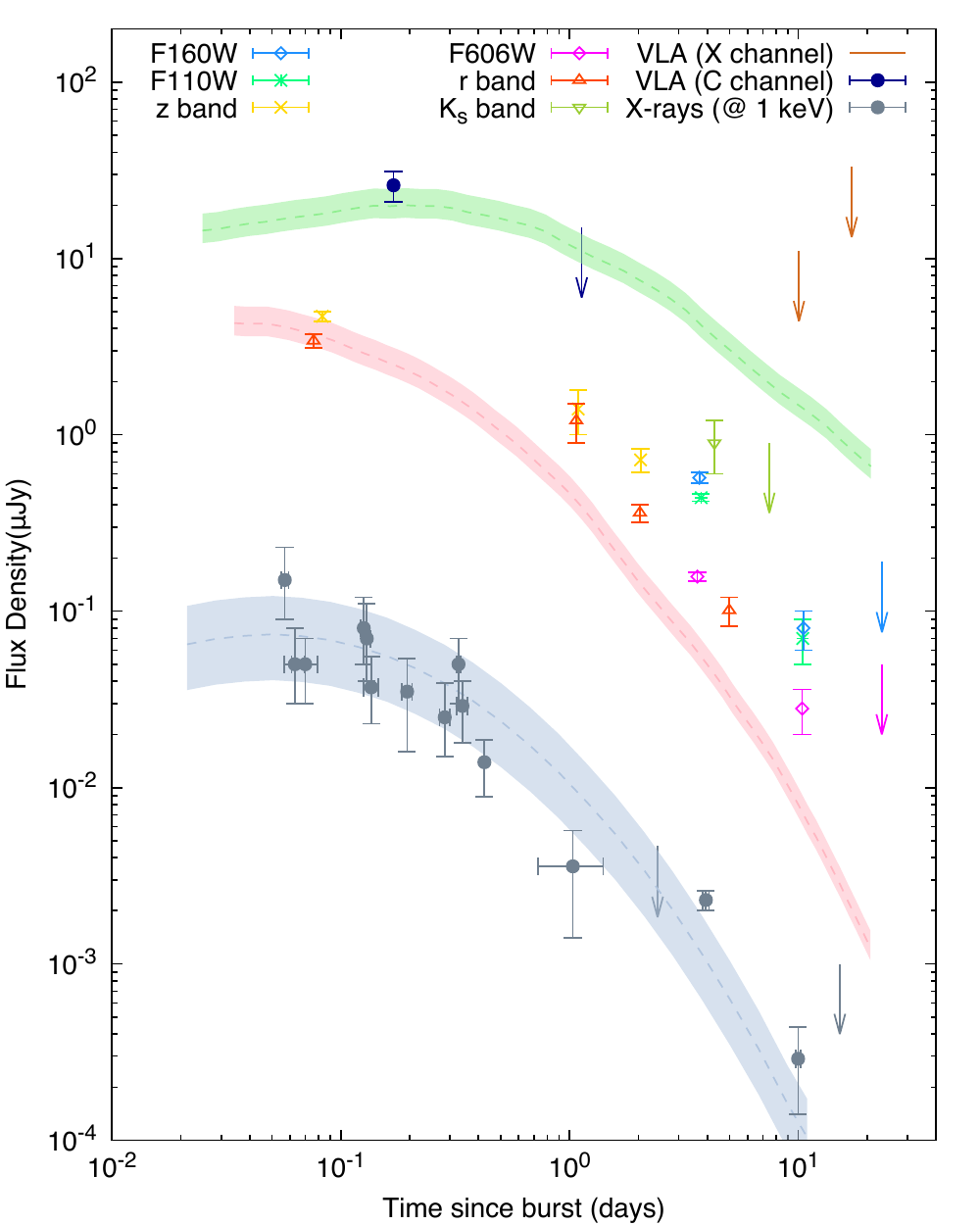}}
\resizebox*{0.5\textwidth}{0.35\textheight}
{\includegraphics{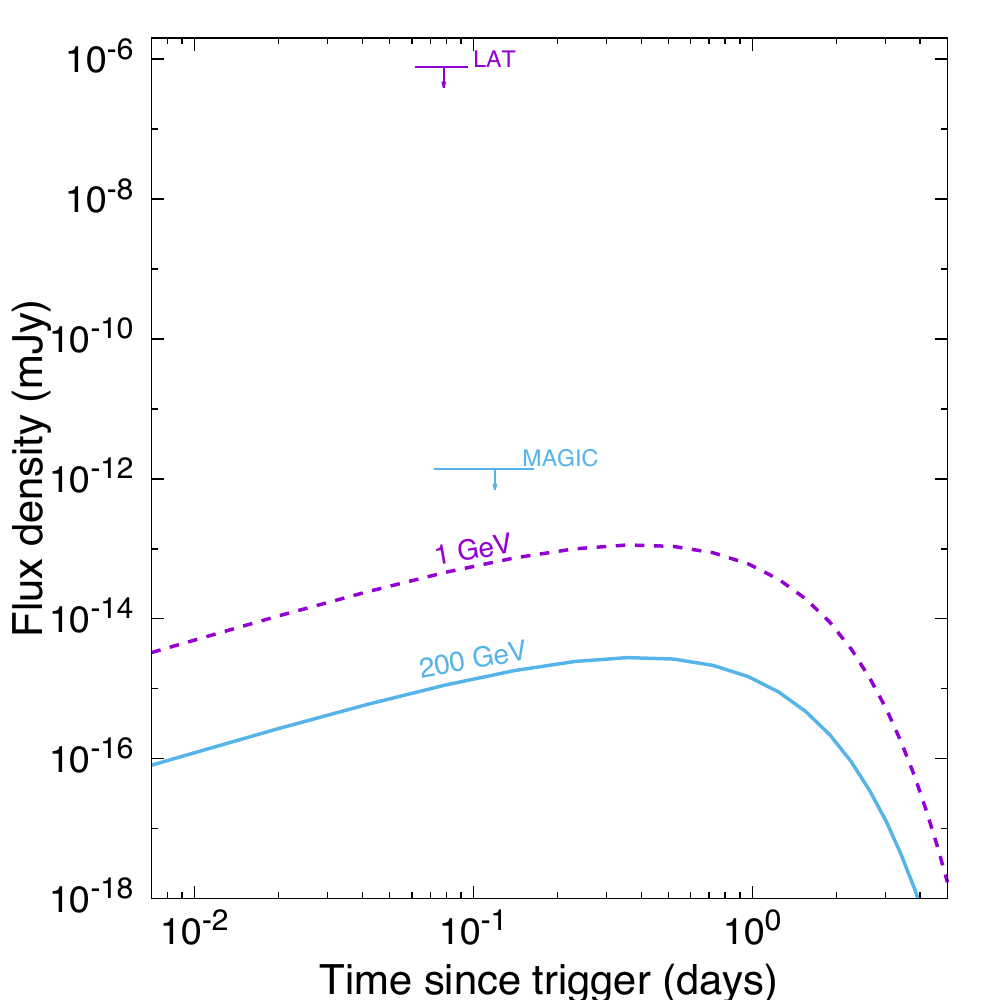}}
}
\caption{Left: The best-fit synchrotron light curves generated when the outflow is decelerated in a uniform  ISM-like medium. These synchrotron light curves are displayed at radio (green), optical (red) and X-ray (gray) energy bands with the data points and upper limits of GRB 160821B.    The best-fit values of the parameters used in our model for X-rays (column 9) are reported in Table \ref{parameters}. Right:  upper limits placed by the Fermi-LAT  and the MAGIC  with the SSC light curves obtained in our model at 1 GeV (purple) and 200 TeV (blue) generated in a uniform ISM-like medium. The effect of the extragalactic background light (EBL) absorption described in \cite{2017A&A...603A..34F} is considered.}
\label{LC_GRB160821B}
\end{figure}

The Swift BAT and Fermi GBM triggered and located  GRB 160821B at 22:29:13  and 22:29:13.33 UT on 2016 August 21, respectively. The Swift XRT  data were obtained using the  public available database at the official Swift web site\footnote{https://swift.gsfc.nasa.gov/cgi-bin/sdc/ql?}. The flux density is extrapolated from 10  keV to 1 keV using the conversion factor introduced in \cite{2010A&A...519A.102E}.   Details of the optical and radio afterglow observations with their upper limits are reported in  \cite{2019MNRAS.489.2104T}.  Fermi LAT searched for  high-energy $\gamma$-ray emission in the 0.1- 300 GeV range and  MAGIC searched for VHE photons above $>500$ GeV from GRB 160821B. In both cases, no photons were detected at the position of this burst and upper limits were derived\footnote{https://pos.sissa.it/312/084/a1.pdf}.\\ 
To obtain the best-fit values of the parameters that fit of the X-ray light curve of GRB 160821B,  once again we use the MCMC simulations.  In this case,  a total of 18200 samples and 8100 tuning steps were performed to describe the entire data.  The best-fit values and the median of the posterior distributions of the parameters are exhibited in Figure \ref{GRB160821B_X-ray}.  The best-fit values of the X-ray band are shown in green color and the medians of the posterior distributions are reported in Table \ref{parameters} (column 9).\\ 
Figure \ref{LC_GRB160821B} shows the multi-wavelength observations of GRB 160821B from 0.2 to 5 days, after the GBM trigger.   The upper limit collected with the Fermi-LAT was obtained from the online data repository\footnote{http://fermi.gsfc.nasa.gov/ssc/data} and the upper limit derived with  the MAGIC observatory is publicly available.\footnote{https://pos.sissa.it/312/084/a1.pdf}  The left-hand panel shows the best-fit light curves obtained using the synchrotron forward-shock model that evolves  in a homogeneous density.  These light curves are shown at the radio, optical and X-ray bands. The radio, optical and X-ray light curves are displayed at  8 GHz, 3 eV and 1 keV, respectively. It is worth noting that although our off-axis model can describe the X-ray and radio observations, it cannot explain the entire optical data. Therefore, the  kilonova-like emission as proposed by \cite{2019MNRAS.489.2104T} and \cite{2019ApJ...883...48L} has to be required.  In our analysis we did not consider the 5 GHz radio afterglow observations that were described with a contribution from a reverse shock \citep{2019ApJ...883...48L}.   The best-fit values of the circumburst density,  the spectral index of the electron population,  the microphysical parameters and the viewing angle are similar to those recently reported in \cite{2019MNRAS.489.2104T} and \cite{2019ApJ...883...48L}.   Given the observed flux of the short peak detected by Fermi GBM  \citep{2016GCN.19843....1S} and  the long-lasting multiwavelength emission,  it can be observed that the short peak  is weaker than the long-lasting multiwavelength emission.  We conclude that the bright peak and the long-lasting afterglow emission were viewed nearly off-axis.  The best-fit value  of the circumburst medium obtained  suggests that the progenitor of GRB 160821B, like other short bursts, exploded in a low density environment. On the other hand,  \cite{2017ApJ...835..181L} assumed that the central engine of GRB 160821 was  a new born supra-massive magnetar and then could interpret this burst in the framework of  the jet radiation and the spin-down  of the pulsar wind.\\
The right-hand figure shows the upper limits derived with the Fermi-LAT and MAGIC observatories with the SSC light curves derived in this work.  We obtain the VHE $\gamma$-ray light curves at 1 GeV (purple) and 200 GeV (blue)  using the values found after describing  the X-ray and optical light curves of GRB 160821B.  The effect of the extragalactic background light (EBL) absorption described in \cite{2017A&A...603A..34F} is used.   With the best-fit values found for this burst, the break energy derived in the KN regime is 486 GeV which is above the VHE upper limits set by Fermi-LAT and MAGIC observatories.    This panel  shows that the SSC flux is consistent with LAT and MAGIC upper limits. Therefore, the SSC model as well as the values used to fit the delayed non-thermal emission are in accordance with the observations.  
\subsection{GRB 170817A}
\begin{figure}[h!]
{ \centering
\resizebox*{0.7\textwidth}{0.4\textheight}
{\includegraphics{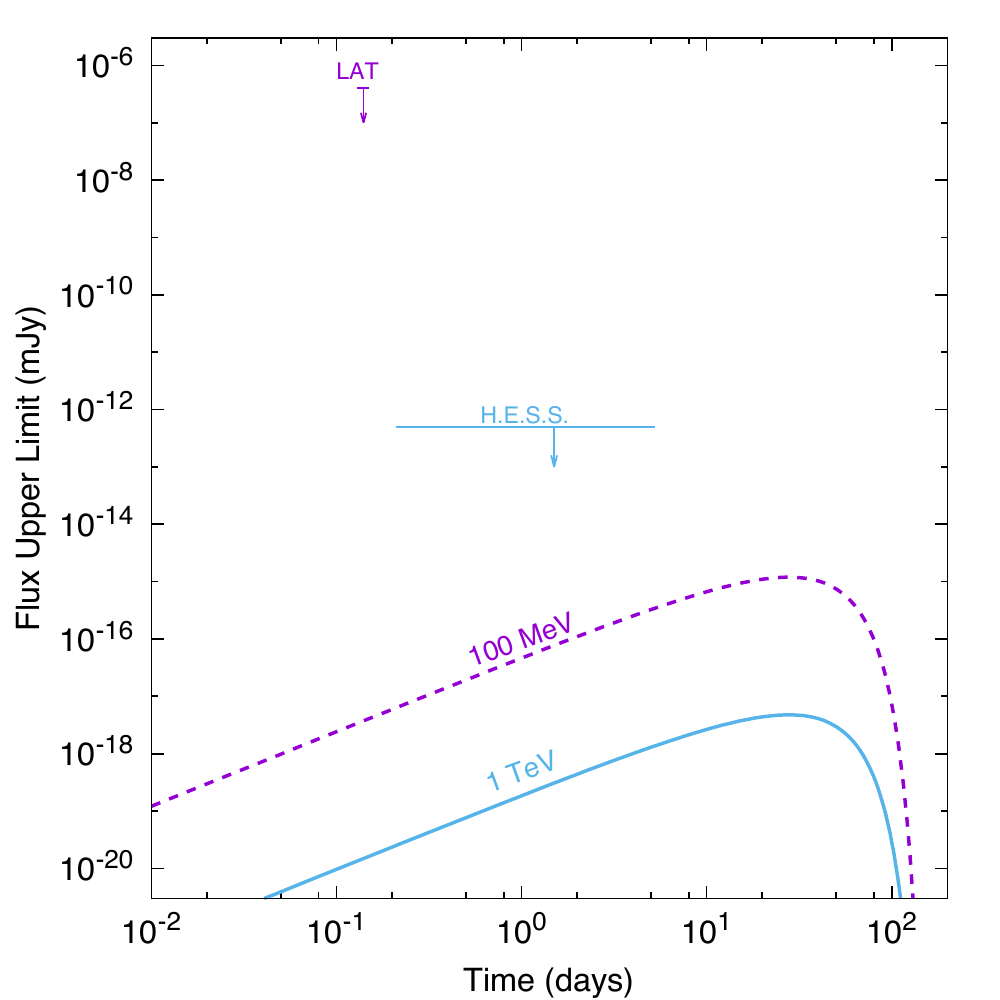}}
}
\caption{Upper limits placed by the Fermi-LAT  and the H.E.S.S. experiment  \citep{2017ApJ...850L..22A} with the SSC light curves obtained in our model at 100 MeV (purple) and 1 TeV (blue) generated in a uniform ISM-like medium. The effect of the extragalactic background light (EBL) absorption described in \cite{2017A&A...603A..34F} is considered.  We use  the best-fit values found with our MCMC code after modelling the X-ray, optical and radio data points of GRB 170817A \cite[see Table 5 in][]{2019ApJ...871..123F}.}
\label{LC_GRB170817A}
\end{figure}

\begin{figure}
	{ \centering
		\resizebox*{\textwidth}{0.7\textheight}
		{\includegraphics[angle=-90]{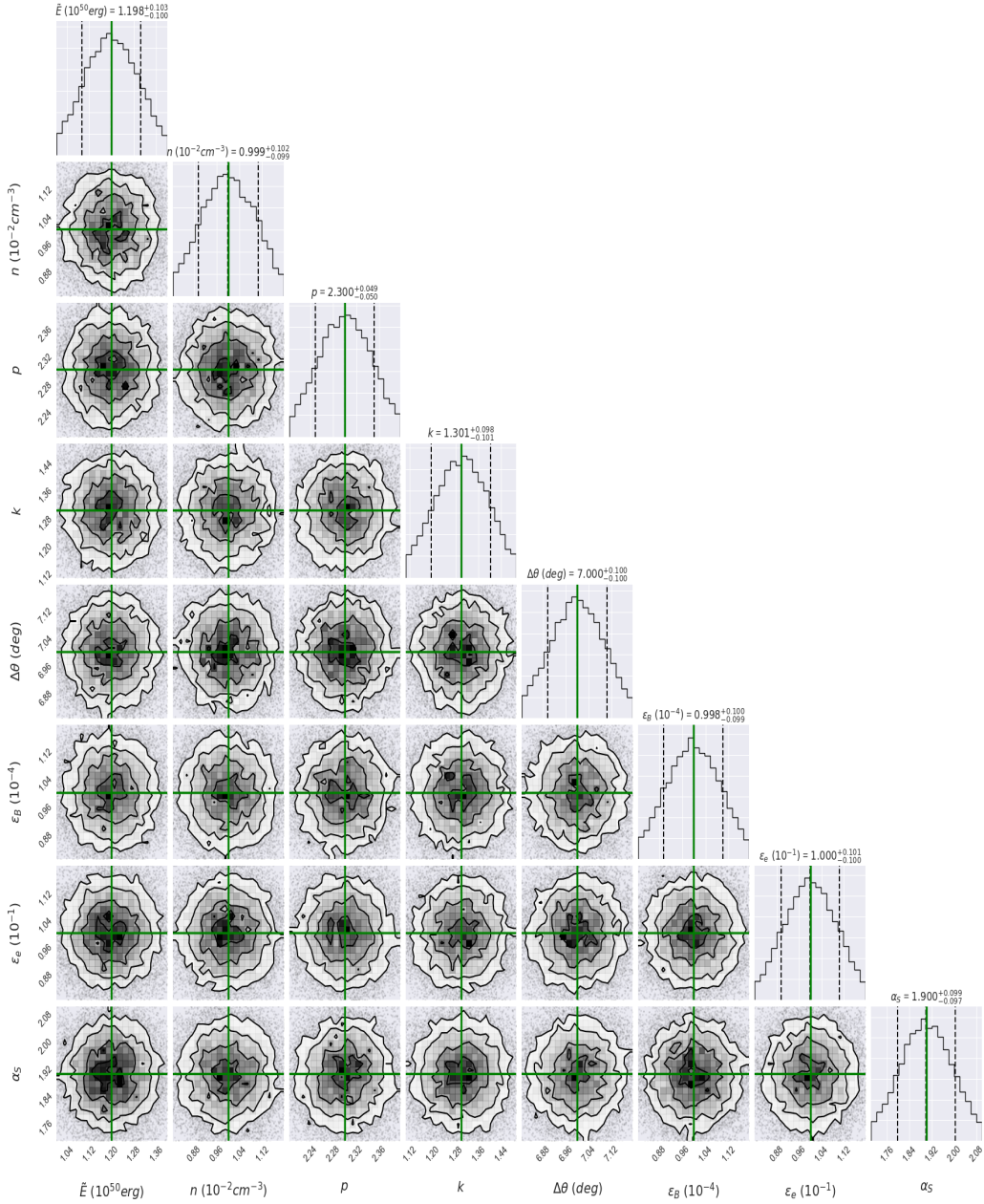}}
	}
	\caption{Corner plot demonstrating the results obtained from the MCMC simulation for our parameter set.  Fit results for the optical light curve  of GRB 080503 using  the synchrotron forward shock model produced by a decelerated jet  in a homogeneous medium viewed off-axis.  Labels above the 1-D KDE plot indicate the quantiles chosen for each parameter.  The best-fit value is shown in green.  Values are reported in Table \ref{parameters} (Column 2).}
	\label{GRB080305_optical}
\end{figure}

\begin{figure}
	{ \centering
		\resizebox*{\textwidth}{0.7\textheight}
		{\includegraphics[angle=-90]{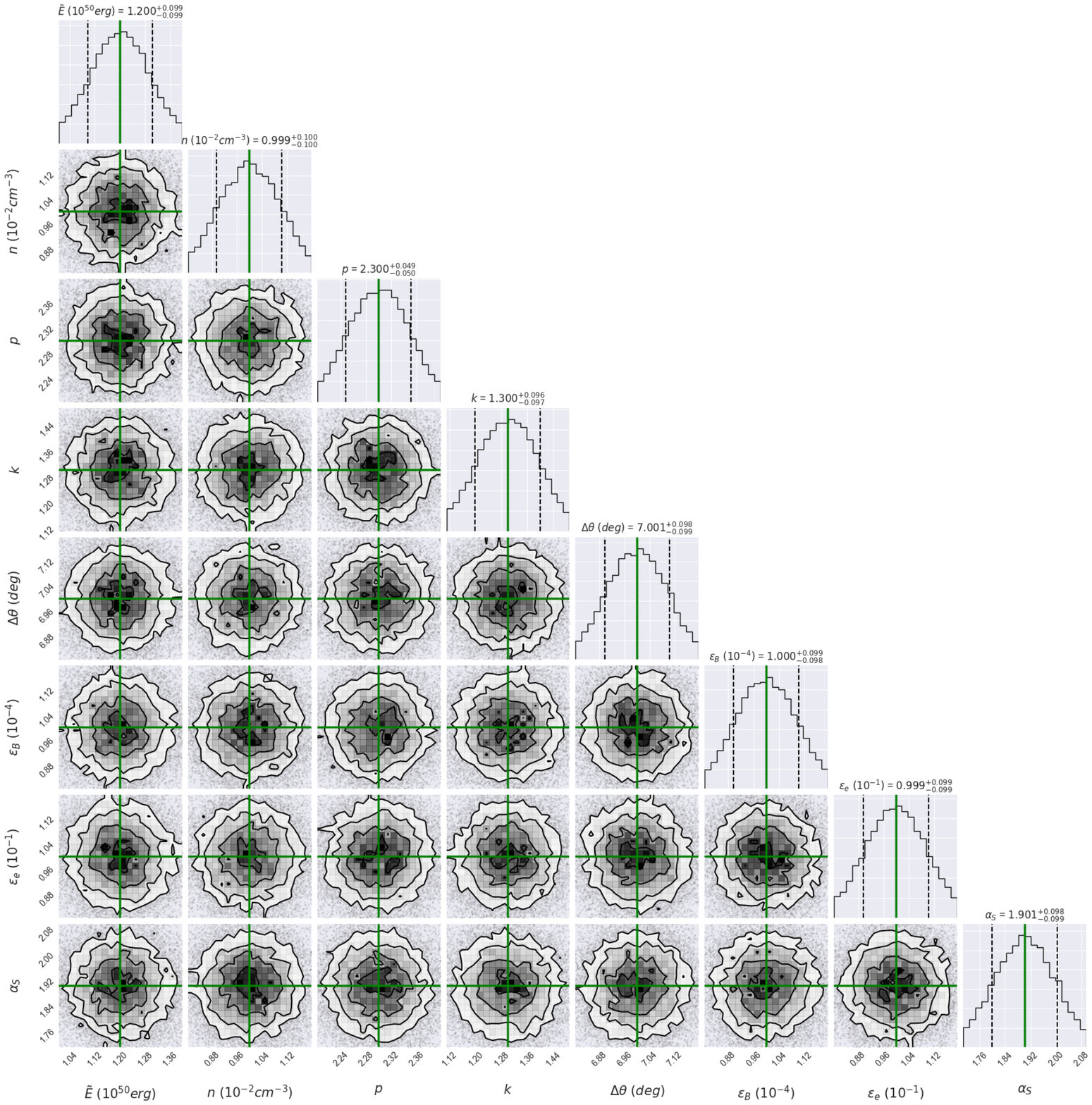}}
	}
	\caption{Same as Fig. \ref{GRB080305_optical},  but it shows the fit results for the X-ray light curve of GRB 080503.  Values are reported in Table \ref{parameters} (Column 3).}
	\label{GRB080305_X-ray}
\end{figure}

\begin{figure}
	{ \centering
		\resizebox*{\textwidth}{0.7\textheight}
		{\includegraphics[angle=-90]{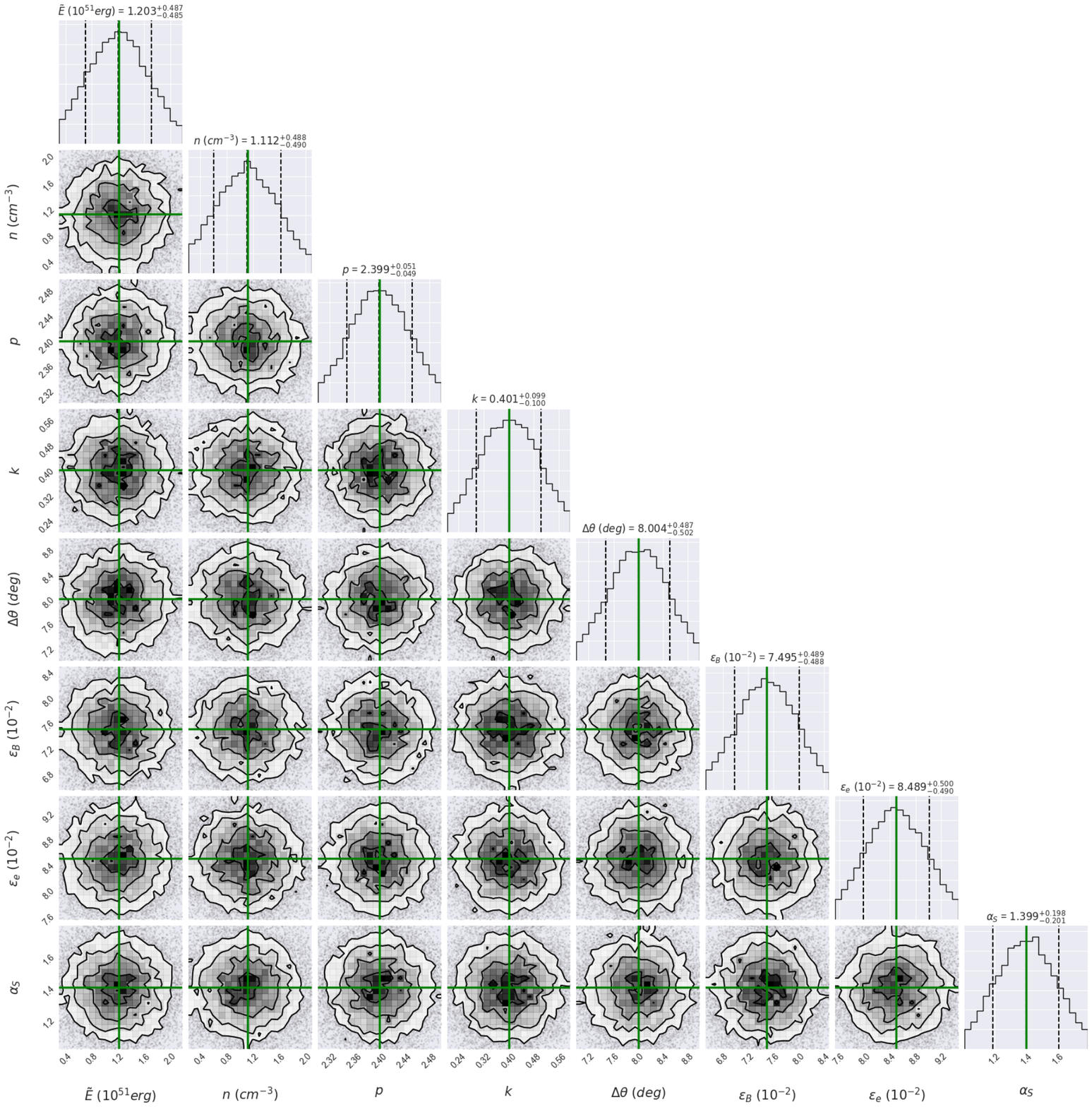}}
	}
	\caption{Same as Fig. \ref{GRB080305_optical},  but it shows the fit results for the radio light curve of GRB 140903A.  Values are reported in Table \ref{parameters} (Column 4).}
	\label{GRB140903A_radio}
\end{figure}

\begin{figure}
	{ \centering
		\resizebox*{\textwidth}{0.7\textheight}
		{\includegraphics[angle=-90]{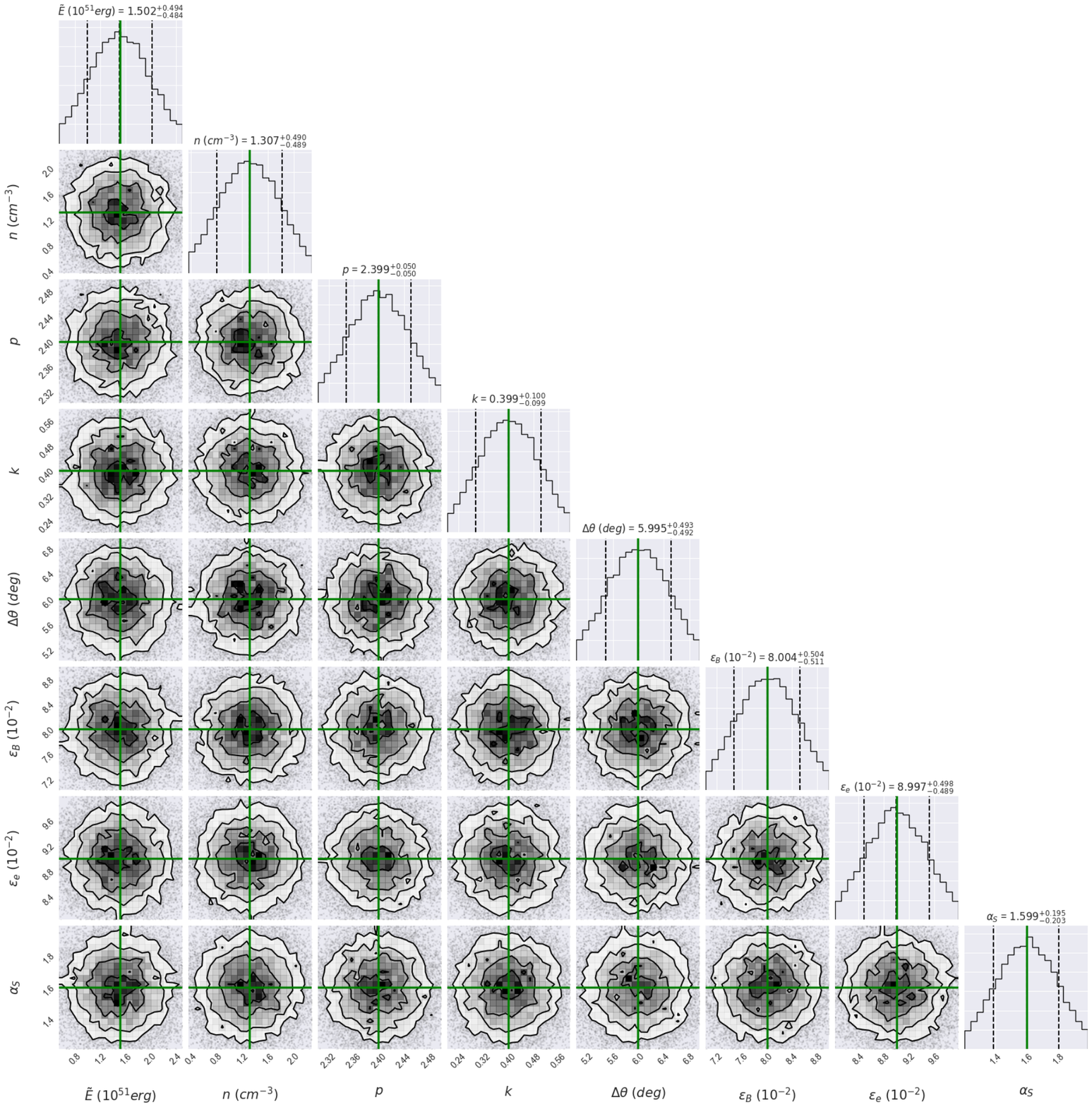}}
	}
	\caption{Same as Fig. \ref{GRB080305_optical},  but it shows the fit results for the optical light curve of GRB 140903A.  Values are reported in Table \ref{parameters} (Column 5).}
	\label{GRB140903A_optical}
\end{figure}

\begin{figure}
	{ \centering
		\resizebox*{\textwidth}{0.7\textheight}
		{\includegraphics[angle=-90]{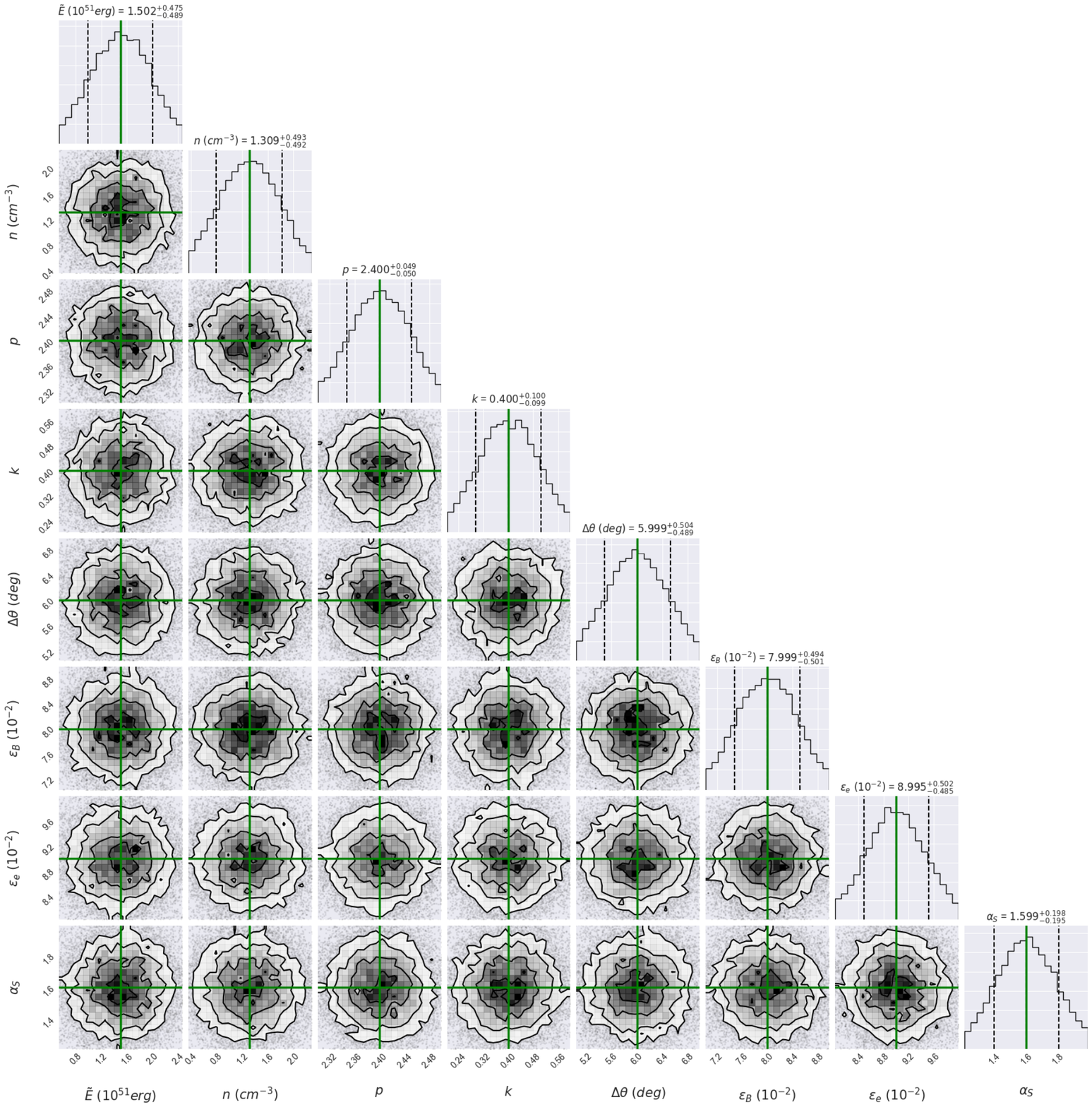}}
	}
	\caption{Same as Fig. \ref{GRB080305_optical},  but it shows the fit results for the X-ray light curve of GRB 140903A.  Values are reported in Table \ref{parameters} (Column 6).}
	\label{GRB140903A_X-ray}
\end{figure}

\begin{figure}
	{ \centering
		\resizebox*{\textwidth}{0.7\textheight}
		{\includegraphics[angle=-90]{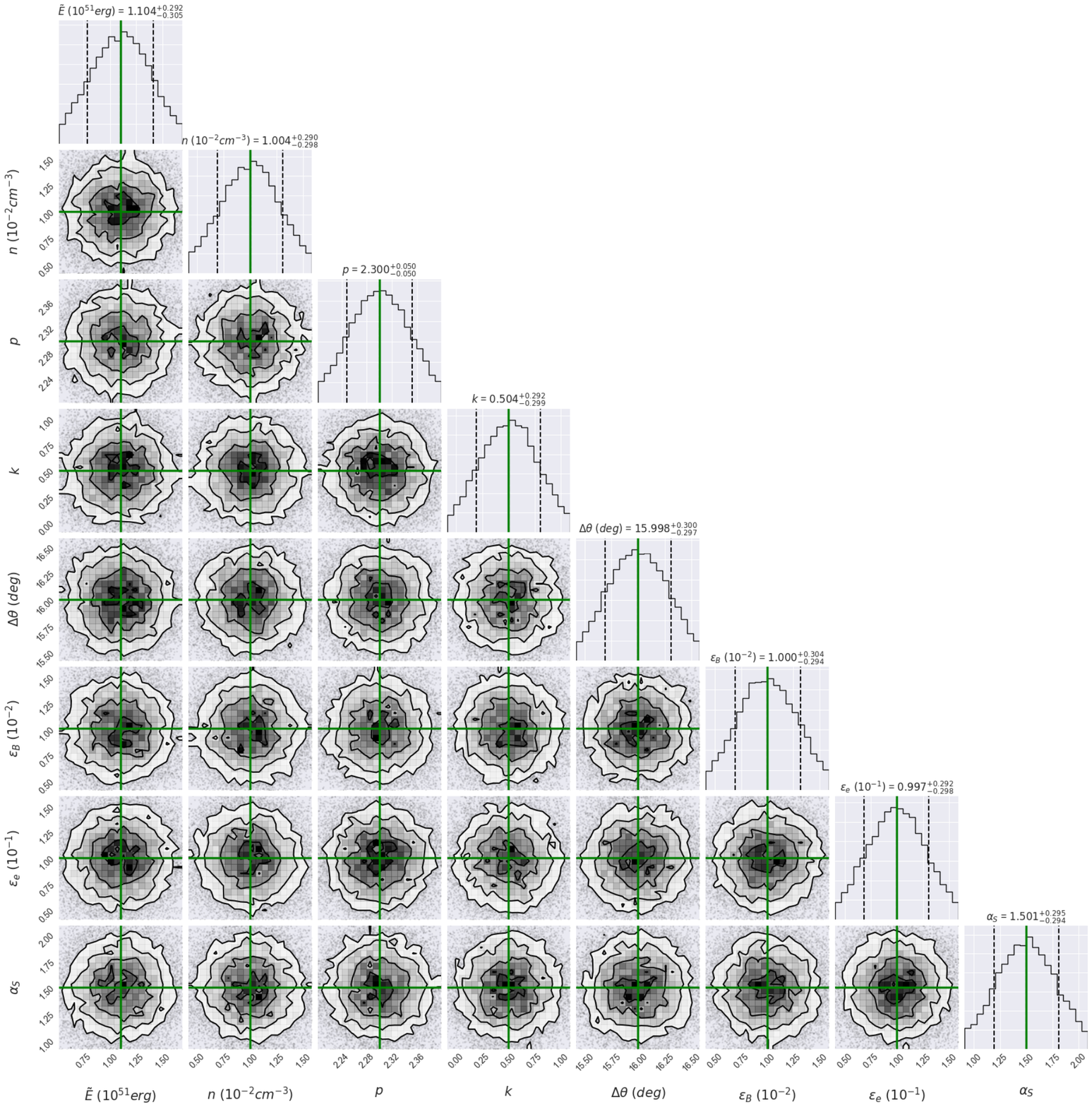}}
	}
	\caption{Same as Fig. \ref{GRB080305_optical},  but it shows the fit results for the X-ray light curve of GRB 150101B.  Values are reported in Table \ref{parameters} (Column 7).}
	\label{GRB150101B_X-ray}
\end{figure}

%
%
\cite{2019ApJ...871..123F} described in detail the multi-wavelength data collected for this event. Here we use the SSC model with the parameters found by the authors and the VHE $\gamma$-ray upper limits.   The Large Area Telescope (Fermi-LAT) and  The High Energy Stereoscopic System  H.E.S.S. Imaging Air Cherenkov Telescope searched for VHE $\gamma$-ray emission from the GW170817 event \citep{2041-8205-848-2-L12, 2017ApJ...850L..22A}.  GW170817A  was in the field of view  of Fermi-LAT $\sim$ 1000 s after the merger trigger. No significant excess was detected at the position of GW170817 and upper limits were derived \citep{2017ApJ...850L..22A}.  Observations with the H.E.S.S. $\gamma$-ray telescope were obtained in two occasions.  The first observation was obtained 5.3 h after the GW trigger.  During the second epoch the HESS observatory covered timescales from 0.22 to 5.2 days and an energy range from 270 GeV to 8.55 TeV.   Although no statistically significant excess of counts was found by this TeV observatory, constraining upper limits were derived.\\
Figure \ref{LC_GRB170817A} shows the upper limits placed with the Fermi-LAT and H.E.S.S. observatories and the corresponding SSC light curves derived in this work.  We derive the VHE $\gamma$-ray light curves at 100 MeV (purple) and 1 TeV (blue)  using the values found by \cite{2019ApJ...871..123F} after describing  the X-ray, optical and radio light curves of GRB 170817A.  The effect of the extragalactic background light (EBL) absorption described in \cite{2017A&A...603A..34F} is used. With the best-fit values found for GRB 170817A, the break energy derived in the KN regime is 2.6 TeV which is above the VHE upper limits set by Fermi-LAT and H.E.S.S. observatories.    As shown in this figure, the SSC flux is too low to be detected by LAT and H.E.S.S.  observatories. Therefore, the SSC model as well as the values used to fit the delayed non-thermal emission are in accordance with the observations reported by the GeV-TeV $\gamma$-ray observatories.
%
%
\section{Conclusions}
%
Several studies have modelled the evolution of the afterglow requiring  the synchrotron emission generated by the deceleration of a relativistic jet seen off-axis.  In particular, some of them have discussed the afterglow, opening angle, jet geometry  and orientation  \cite[e.g., see][]{2017MNRAS.471.1652L, 2018ApJ...857..128J, 2018MNRAS.481.2581L, 2018PhRvL.120x1103L}.   In this paper,  we have extended the analytical scenario shown in \cite{2019ApJ...871..123F} by deriving, for a more general case,  the SSC and synchrotron forward-shock light curves when this outflow is decelerated in a homogeneous and wind-like circumburst medium  in the fully adiabatic regime.  In the particular case of $\delta=0$, the SSC and synchrotron light curves derived in a homogeneous and wind-like medium are recovered \citep{1998ApJ...497L..17S, 2000ApJ...536..195C, 2001ApJ...548..787S, 1999ApJ...519L..17S}. We have computed the light curves considering the values of observables and parameters in the typical ranges:  $\tilde {E}=10^{50} - 10^{52}\,{\rm erg}$, $n=10^{-4} - 10^{-3}\,{\rm cm^{-3}}$,  $A_\star=1- 10^4$,  $\epsilon_B=10^{-4} - 10^{-1}$, $\Delta \theta=15^\circ - 30^\circ$ and $\alpha_s=1.1 - 2.1$ for the values of $\epsilon_e=0.1$, $p=2.2$ and $D=100 {\rm Mpc}$.\\
%
%
\cite{2018ApJ...863L..34B} analyzed the prompt phase of GRB 150101B.  These authors argued that the prompt emission was formed by a two-component structure; a short hard spike followed by a longer soft tail. Authors concluded that the cocoon shock breakout models disfavor the description of the two-component structure in this light curve. They  derived the conditions  for radius of acceleration to take place below the photospheric radius, assuming a wind-like medium in the vicinity of the NS merger.  These authors found that the values of the density parameter and mass density were $A\gtrsim 4.5\times 10^{35}\,{\rm g\,cm^{-1}}$  and $\rho\gtrsim 10^{-2}\,{\rm g\, cm^{-3}}$, respectively.   In the model proposed in this paper, we showed that the flux emitted from synchrotron forward-shock emission in a wind-like medium  is in the range of the Fermi GBM  for values of $A\sim10^{39}\,{\rm g\,cm^{-1}}$ and $\rho\sim 1\,{\rm g\, cm^{-3}}$ which agree with those derived in \cite{2018ApJ...863L..34B} and \cite{2013ApJ...773...78B}. If this is the case,  a transition phase from wind-like medium to homogeneous medium is expected as indicated in \cite{2017ApJ...848...15F}.\\
%
%
%
In particular, we have analyzed GRB 080503,  GRB 140903A, GRB 150101B,  GRB 160821B and GRB 170817A.    For  GRB 080503,  GRB 140903A, GRB 150101B and  GRB 160821B we have shown that the origin of the delayed and long-lasting afterglow emission could be interpreted by a similar scenario to the one used to describe GRB 170817A; the  radio, optical and  X-ray light curves with the upper limits through the synchrotron forward-shock model \citep[e.g., see][]{2018MNRAS.478L..18T, 2017Natur.551...71T, 2018PhRvL.120x1103L}.  The non-thermal radio, optical and X-ray fluxes with the upper limits are consistent with the synchrotron forward-shock model in a homogeneous circumburst medium, indicating that the multiwavelength observations were generated by the same power laws and  the  peak times are consistent with the fact that the beaming cone of the  radiation reaches our line of sight. For GRB 160821B, we show additionally that the proposed scenario agrees with the VHE $\gamma$-ray upper limits derived by the TeV $\gamma$-ray observatories. The SSC fluxes are 4 - 8 orders of magnitude less than the high-energy upper limits. For GRB 170817A, the gamma-ray spike and the delayed non-thermal emission was described in \cite{2019ApJ...871..123F}.  Here, we show that the proposed scenario agrees with the VHE $\gamma$-ray upper limits derived by the TeV $\gamma$-ray observatories. The SSC fluxes are 8 - 10 orders of magnitude less than the high-energy upper limits.    It is worth emphasizing that in GRB 080503, GRB 140903A, GRB 160821B and GRB 170817A, the synchrotron forward-shock radiation emitted from a homogeneous medium was favored over the radiation emitted from a stratified stellar-wind medium. For GRB 150101B, the emission of synchrotron forward-shock radiation emitted from both a wind or a homogeneous medium is consistent with the X-ray data and optical upper limits.  In the case of the stratified wind-like medium, our results are consistent with those reported by the GBM collaboration after the description of the short and hard gamma-ray peak.  Based on the parameter values found using our model, we point out that:\\
i) The values of the homogeneous medium required to describe the non-thermal long-lasting afterglow emission suggest that the progenitor of these bursts  exploded in a very low density environment.  These values are  in agreement   with the larger offsets of sGRBs compared with lGRBs.\\
ii) The values of the spectral indexes of the electron populations are consistent with the typical values reported when relativistic electrons accelerated in the forward shocks are cooled down by synchrotron radiation \citep[see, e.g.][]{2015PhR...561....1K, 2017ApJ...848...94F, 2019ApJ...872..118B,  2019arXiv190405987B}.  It reaffirms that the long-lasting afterglow emission was originated in the GRB afterglow.\\
iii) Assuming a value in the range of $4^\circ - 6^\circ$  for the jet opening angle for these bursts, the viewing angles become $1^\circ \lesssim \theta_{\rm obs}\lesssim 10^\circ$.   Given the observed fluxes of the hard and short spikes and  the long-lasting afterglow emissions, the spike components are fainter than  the long-lasting afterglow components.  The fact that the total energy of the delayed non-thermal emission can exceed that of the hard spikes by a large factor is a problem for the NS merger scenario which is limited to some seconds by the viscous timescale \citep[see,  e.g.][]{2004ApJ...608L...5L}. However,  it could be reconciled with the merger scenario as proposed in our model where  the hard spikes focused in a collimated jet are viewed nearly off-axis whereas the long-lasting afterglow emissions are more widely beamed.\\ 
iv) The derived values  of the kinetic energies $\sim 10^{51-52}\,{\rm erg}$  suggest that pair annihilation of neutrinos and anti-neutrinos is a possible mechanism to provide the energy budget $L_{\nu\bar\nu}\lesssim 10^{51} {\rm erg s^{-1}}$. This result agrees with numerical simulation of merging NS-NS or NS-BH systems.\\  
v) The VHE upper limits set by Fermi LAT, MAGIC and H.E.S.S. observatories are below the SSC energy break derived in the KN regime. This result indicates that the SSC break energy is not drastically attenuated, which encourages us to keep observing these events in VHEs.\\
The multi-wavelength light curves indicate that  GRB 080503, GRB 140903A,  GRB 150101B, GRB 160821B and GRB 170817A  originated from the same kind of progenitors, despite their diversity.  We might argue that the short bursts detected by  the BAT and GBM instruments without their corresponding emissions in other electromagnetic bands were too faint during the first second to be detected and followed up.  
%
%
%
\\
\\
%
%
\\
\acknowledgements
 
We  thank  W. Lee and  E. Ramirez-Ruiz  for  useful  discussions.   NF  acknowledges financial support from UNAM-DGAPA-PAPIIT through  grant  IA102019.  FDC  thanks the  UNAM-PAPIIT  grants  AG100820  and  support  from  the Miztli-UNAM  supercomputer  (project  LANCAD-UNAM-DGTIC-281). PV  thanks  Fermi  grants  NNM11AA01A and 80NSSC17K0750,  and partial support from OTKA NN111016 grant. RBD acknowledges support from the National Science  Foundation  under  Grant  1816694. 
%
%
%
%
%

%
%
\newpage
\newpage
\newpage


\clearpage

\clearpage

\begin{figure}
	{ \centering
		\resizebox*{\textwidth}{0.7\textheight}
		{\includegraphics[angle=-90]{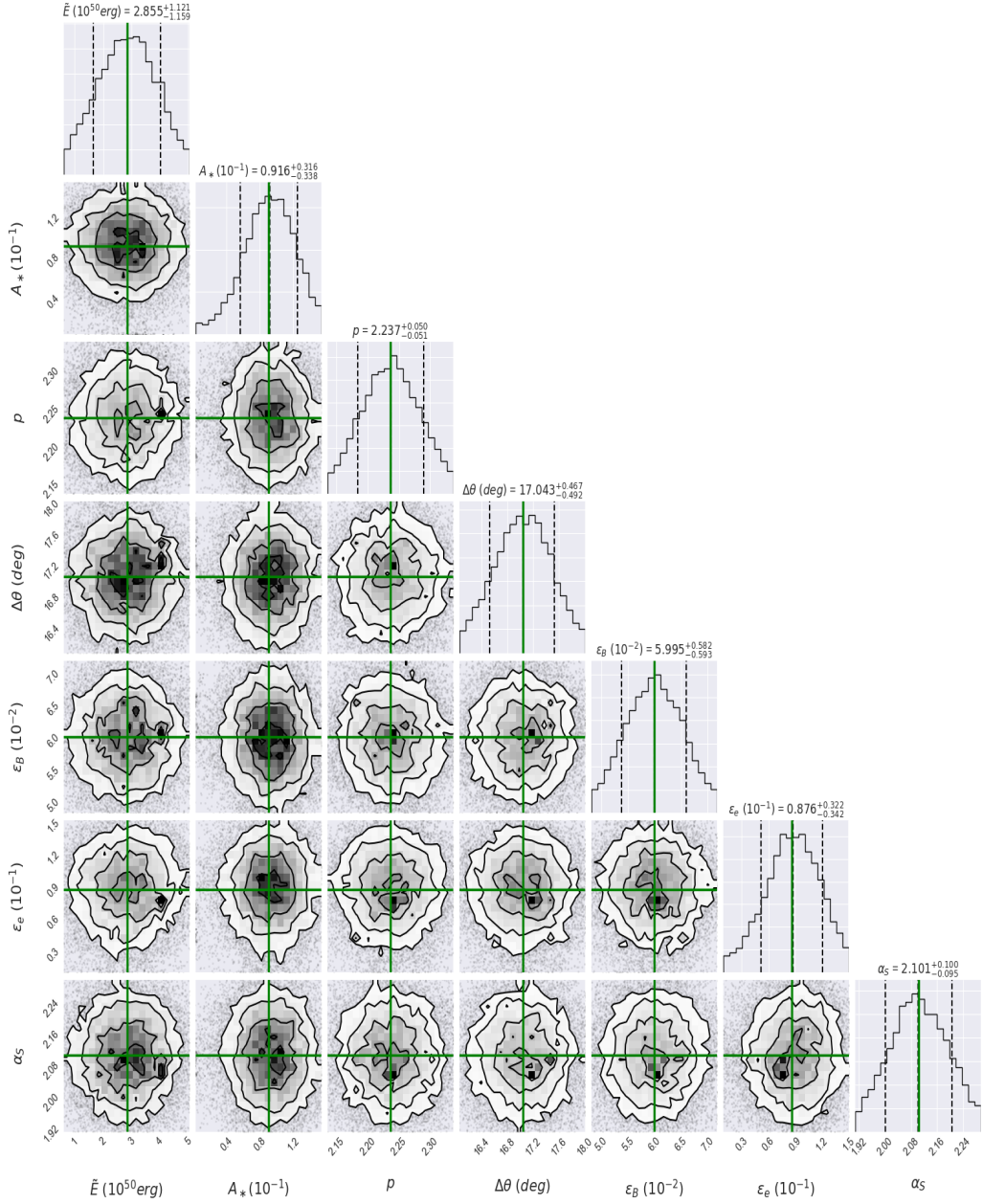}}
	}
	\caption{Same as Fig. \ref{GRB150101B_X-ray},  but it shows the fit results for the X-ray light curve of GRB 150101B  using the synchrotron forward shock model produced by a decelerated jet viewed off-axis in a wind-like medium.  Values are reported in Table \ref{parameters} (Column 8).}
	\label{GRB150101B_X-ray_wind}
\end{figure}

\begin{figure}
	{ \centering
		\resizebox*{\textwidth}{0.7\textheight}
		{\includegraphics[angle=-90]{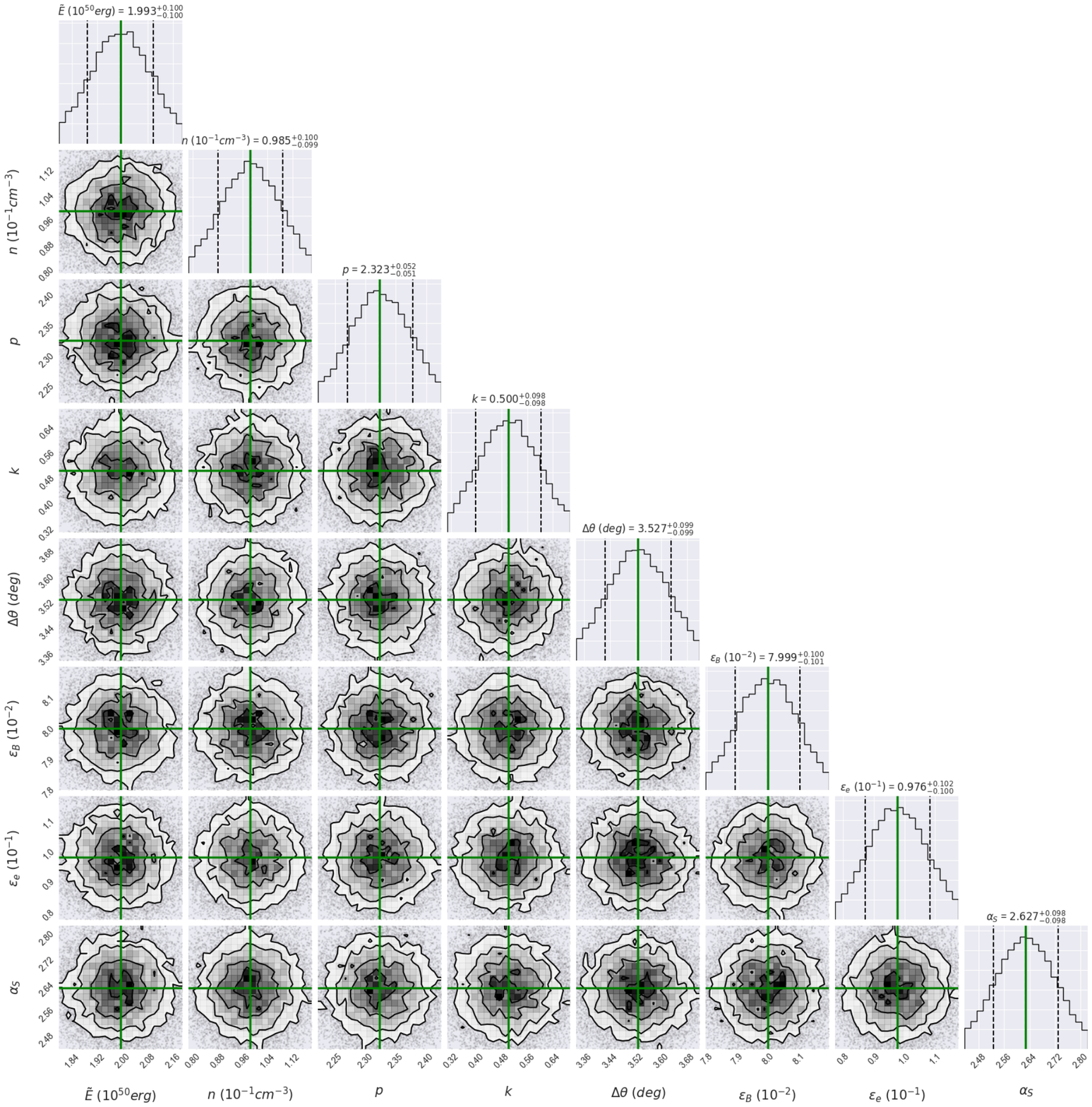}}
	}
	\caption{Same as Fig. \ref{GRB080305_optical},  but it shows the fit results for the X-ray light curve of GRB 160821B.  Values are reported in Table \ref{parameters} (Column 11).}
	\label{GRB160821B_X-ray}
\end{figure}

%

\begin{sidewaystable}[h!]
\centering \renewcommand{\arraystretch}{2}\addtolength{\tabcolsep}{-1pt}
\caption{The median and symmetrical quantiles (0.15, 0.5, 0.85) are reported after the description of the X-ray, optical and radio observations for GRB 080503,  GRB 140903A, GRB 150101B and GRB 160821B. These values are obtained using the theoretical model  and the MCMC simulations.}\label{parameters}
\begin{tabular}{  l  |   c  |  c  | c  |  c }
\hline
\hline


{\large   Parameters}	&      {\large  GRB 080503}  & 	{\large  GRB 140903A}	  & {\large  GRB 150101B} & {\large  GRB 160821B} \\
        &  {\normalsize Optical}  	\hspace{0.9cm}	{\normalsize X-ray} & 		 {\normalsize Radio}  \hspace {1cm } {\normalsize Optical}  	\hspace{1cm}	{\normalsize X-ray} &  {\normalsize X-ray}	& 	{\normalsize X-ray} \\ 

        &  {\normalsize }  	\hspace{0.9cm}	{\normalsize} & 		 {\normalsize}  \hspace {1cm } {\normalsize }  	\hspace{1cm}	{\normalsize} &  {\normalsize (Wind) \hspace{1.7cm} (ISM)}	& 	{\normalsize} \\

\hline \hline
&  &  &\\
\small{$\tilde{E}\, (10^{50}\,{\rm erg})$}	\hspace{0.2cm}&  \small{$1.19^{+0.10}_{-0.10}$}	\hspace{1cm}   \small{$1.20^{+0.09}_{-0.09}$}	&  \small{$(1.20^{+0.48}_{-0.48}$} \hspace{0.3cm}   \small{$1.50^{+0.49}_{-0.48}$}	\hspace{0.3cm}    \small{$1.50^{+0.47}_{-0.48})\times10^1$} &    \small{$2.85^{+1.12}_{-1.15}\times10^1$} \hspace{1cm} \small{$1.10^{+0.29}_{-0.30}\times10^1$} 	&   \small{$1.99^{+0.10}_{-0.10}$}	 \\

\small{${\rm n}\,\, (10^{-1}\,{\rm  cm^{-3}}$) }	\hspace{0.2cm}&  \small{($0.99^{+0.10}_{-0.09}$}	\hspace{0.2cm}   \small{$0.99^{+0.10}_{-0.10})\times10^{-1}$}	&  \small{$(1.11^{+0.48}_{-0.49}$} \hspace{0.3cm}   \small{$1.307^{+0.49}_{-0.48}$}	\hspace{0.3cm}    \small{$1.30^{+0.49}_{-0.49})\times 10 $} & \hspace{0.65cm}  \small{$-$} \hspace{1.65cm}  \small{$1.00^{+0.29}_{-0.29}\times10^{-1}$}	&   \small{$0.98^{+0.10}_{-0.09}$}	 \\

\small{${\rm A_{\star}}\,\, (10^{-1}\,$) }	\hspace{0.2cm}&  \small{$$}	\hspace{0.2cm}   \small{$$}	&  \small{$$} \hspace{0.3cm}   \small{$$}	\hspace{0.3cm}    \small{$ $} &  \small{$0.91^{+0.31}_{-0.33}\times10^{-1}$} \hspace{1.25cm}  \small{$-$} \hspace{1.cm}	&   \small{$$}	 \\

\small{${\rm p}$ }	\hspace{0.2cm}&  \small{$2.30^{+0.04}_{-0.05}$}	\hspace{1cm}   \small{$2.30^{+0.04}_{-0.05}$}	&  \small{$2.39^{+0.05}_{-0.04}$} \hspace{0.6cm}   \small{$2.39^{+0.10}_{-0.09}$}	\hspace{0.6cm}    \small{$2.40^{+0.04}_{-0.05}$} & \small{$2.23^{+0.05}_{-0.05}$}    \hspace{1cm}   \small{$2.30^{+0.05}_{-0.05}$}	&   \small{$2.32^{+0.05}_{-0.05}$}	 \\

\small{${\rm k}$ }	\hspace{0.2cm}&  \small{$1.30^{+0.09}_{-0.10}$}	\hspace{1cm}   \small{$1.30^{+0.09}_{-0.09}$}	&  \small{$0.40^{+0.09}_{-0.10}$} \hspace{0.6cm}   \small{$0.39^{+0.10}_{-0.09}$}	\hspace{0.6cm}    \small{$0.40^{+0.10}_{-0.09}$} & \small{$-$} \hspace{1.75cm} 	   \small{$0.50^{+0.29}_{-0.29}$}  & \small{$0.50^{+0.09}_{-0.09}$}	 \\

\small{$\Delta \theta$\,({\rm deg})}	\hspace{0.2cm}&  \small{$7.00^{+0.10}_{-0.10}$}	\hspace{1cm}   \small{$7.00^{+0.09}_{-0.09}$}	&  \small{$8.00^{+0.48}_{-0.50}$} \hspace{0.6cm}   \small{$5.99^{+0.49}_{-0.49}$}	\hspace{0.6cm}    \small{$5.99^{+0.50}_{-0.48}$} &  \small{$17.04^{+0.46}_{-0.49}$} \hspace{1cm}   \small{$15.99^{+0.30}_{-0.29}$}	& \small{$3.52^{+0.09}_{-0.09}$}	 \\

\small{$\varepsilon_{\rm B}\,\,(10^{-3})\,$}	\hspace{0.2cm}&  \small{$(0.99^{+0.10}_{-0.09}$}	\hspace{0.2cm}   \small{$1.00^{+0.09}_{-0.09})\times10^{-1}$}	&  \small{$(7.49^{+0.48}_{-0.48}$} \hspace{0.3cm}   \small{$8.00^{+0.50}_{-0.51}$}	\hspace{0.3cm}    \small{$7.99^{+0.49}_{-0.50})\times10$} &  \small{$5.99^{+0.58}_{-0.59}\times10$} \hspace{1cm} \small{$1.00^{+0.30}_{-0.29}\times10$}	&  \small{$0.79^{+0.10}_{-0.10}$}	 \\

\small{$\varepsilon_{\rm e}\,\,(10^{-1})\,$}	\hspace{0.2cm}&  \small{$1.00^{+0.10}_{-0.10}$}	\hspace{1cm}   \small{$0.99^{+0.09}_{-0.09}$}	&  \small{$(8.48^{+0.50}_{-0.49}$} \hspace{0.3cm}   \small{$8.99^{+0.49}_{-0.48}$}	\hspace{0.3cm}    \small{$8.99^{+0.50}_{-0.48})\times10^{-1}$} &\small{$0.87^{+0.32}_{-0.34}$} \hspace{1cm} \small{$0.99^{+0.29}_{-0.29}$}	&  \small{$0.97^{+0.10}_{-0.10}$}	 \\

\small{$\alpha_{\rm s}\,$}	\hspace{0.2cm}&  \small{$1.90^{+0.09}_{-0.09}$}	\hspace{1cm}   \small{$1.90^{+0.09}_{-0.09}$}	&  \small{$1.39^{+0.19}_{-0.20}$} \hspace{0.6cm}   \small{$1.59^{+0.19}_{-0.20}$}	\hspace{0.6cm}    \small{$1.59^{+0.19}_{-0.19}$} &  \small{$2.10^{+0.10}_{-0.09}$}  \hspace{1cm}  \small{$1.50^{+0.29}_{-0.29}$}	& \small{$2.62^{+0.09}_{-0.09}$}	 \\

\hline
\end{tabular}
\end{sidewaystable}
\end{document}